\documentclass[usenatbib]{mn2e}
\usepackage{graphicx}
\usepackage{rotating}
\usepackage{txfonts}
\usepackage{fixads}
\usepackage[british]{babel}
\usepackage{dcolumn}
\newcolumntype{d}{D{.}{.}{-1}}
\def\dhead#1{\multicolumn{1}{c}{#1}}
\def\twolines#1#2{$\kern-6pt\Big\{ {\textrm{#1\hfill}\atop\textrm{#2\hfill}}$}
\graphicspath{{.}{./FIGS/}}
\title[The faint source population at 15.7~GHz]{The faint source population at 15.7~GHz -- I. The radio properties}

\label{firstpage}

\author[I. H. Whittam et al.]{\parbox{\textwidth}{I. H. Whittam,$^{1}$\thanks{E-mail:
ihw24@mrao.cam.ac.uk} J. M. Riley,$^{1}$ D. A. Green,$^{1}$ M.~J.~Jarvis,$^{2,3}$ I.~Prandoni,$^{4}$ G.~Guglielmino,$^{4,5,6}$ R.~Morganti,$^{6,7}$ H.~J.~A.~R{\"o}ttgering$^{8}$ and M.~A.~Garrett$^{6,8}$}\vspace{0.4cm}\\
   $^{1}$Astrophysics Group, Cavendish Laboratory, 19 J.~J.~Thomson Avenue, Cambridge CB3 0HE\\
   $^{2}$Centre for Astrophysics Research, STRI, University of Hertfordshire, Hatfield, AL10 9AB\\
   $^{3}$Physics Department, University of the Western Cape, Bellville 7535, South Africa\\
   $^{4}$INAF-IRA, via Piero Gobetti 101, 40127, Bologna, Italy\\
   $^{5}$Universit\`a di Bologna, via Ranzani 1, 40127, Bologna, Italy\\
   $^{6}$ASTRON, P.O. Box 2, 7990 AA Dwingeloo, The Netherlands\\
   $^{7}$Kapteyn Astronomical Institute, University of Groningen, P. O. Box 800, 9700 AV Groningen, The Netherlands \\
   $^{8}$Leiden Observatory, Leiden University, P. O.  Box 9513, NL-2300 RA Leiden, The Netherlands}

\date{Accepted ---; received ---; in original form ---}

\pagerange{\pageref{firstpage}--\pageref{lastpage}}

\pubyear{2012}
\begin{document}

\maketitle

\begin{abstract}
We have studied a sample of 296 faint ($> 0.5$ mJy) radio sources selected from an area of the Tenth Cambridge (10C) survey at 15.7~GHz in the Lockman Hole. By matching this catalogue to several lower frequency surveys (e.g. including a deep GMRT survey at 610~MHz, a WSRT survey at 1.4~GHz, NVSS, FIRST and WENSS) we have investigated the radio spectral properties of the sources in this sample; all but 30 of the 10C sources are matched to one or more of these surveys. We have found a significant increase in the proportion of flat spectrum sources at flux densities below $\approx$1 mJy -- the median spectral index between 15.7~GHz and 610~MHz changes from 0.75 for flux densities greater than 1.5~mJy to 0.08 for flux densities less than 0.8~mJy. This suggests that a population of faint, flat spectrum sources is emerging at flux densities $\lesssim 1 \textrm{ mJy}$.

The spectral index distribution of this sample of sources selected at 15.7~GHz is compared to those of two samples selected at 1.4~GHz from FIRST and NVSS. We find that there is a significant flat spectrum population present in the 10C sample which is missing from the samples selected at 1.4~GHz. The 10C sample is compared to a sample of sources selected from the SKADS Simulated Sky by Wilman et al. and we find that this simulation fails to reproduce the observed spectral index distribution and significantly underpredicts the number of sources in the faintest flux density bin. It is likely that the observed faint, flat spectrum sources are a result of the cores of FRI sources becoming dominant at high frequencies. These results highlight the importance of studying this faint, high frequency population.
\end{abstract}

\begin{keywords}
galaxies: active -- radio continuum: galaxies -- galaxies: starburst
\end{keywords}

\section{Introduction}

The Tenth Cambridge Survey (10C; \citealt{2011MNRAS.415.2699F,2011MNRAS.415.2708D}) at 15.7~GHz is the deepest  high frequency ($\gtrsim 10$ GHz) radio survey to date, complete to 1~mJy in ten different fields covering a total of $\approx 27 \textrm{ deg}^2$. A further $\approx 12 \textrm{ deg}^2$, contained within these fields, is complete to 0.5~mJy. The 10C survey therefore enables us to study the faint source population at 15.7~GHz, a parameter space which has not been explored in any detail. Most studies of the sub-mJy population have focused on lower frequencies, due to the increased time required to survey a field to an equivalent depth at higher frequencies.

One area of particular interest has been the variation with flux density of the spectral index distribution for the samples selected at 15 -- 20 GHz. For example, in the Ninth Cambridge survey (9C; \citealt{2003MNRAS.342..915W}) the proportion of sources with flat or rising spectra decreased as the flux density decreased. The median spectral index between 15.2 and 1.4~GHz ($\alpha^{15.2}_{1.4}$) changed from 0.23 for the highest flux density bin ($S_{15 \rm~GHz} \geq$ 100 mJy) to 0.79 for the lowest flux density bin (5.5 mJy $\geq S_{15 \rm~GHz} >$ 25 mJy). (The convention $S \propto \nu^{-\alpha}$, where $S$ is flux density at frequency $\nu$, is used throughout this work).

The Australia Telescope 20 GHz survey (AT20G; \citealt{2011MNRAS.412..318M}) found a similar variation of spectral index with flux density for a sample with a flux density limit of 40~mJy. The 10C survey \citep{2011MNRAS.415.2708D} enables us to extend these studies to lower flux densities.  Although the study by \citet{2011MNRAS.415.2708D} contains a larger number of limits, it is clear that the fraction of flat spectrum sources increases again as flux density decreases further. 

There have been several attempts to model the high frequency radio sky (\citealt{2008MNRAS.388.1335W}, \citealt{2010MNRAS.405..447W}, \citealt{2005A&A...431..893D}, \citealt{2011A&A...533A..57T}). These simulations  have extrapolated from lower frequency surveys, generally on the assumption that the sources have power-law synchrotron spectra. These models are an increasingly poor fit to the observed sources counts below 10~mJy at 15~GHz. The number of sources is significantly underestimated, indicating that the properties of these sources are not well understood, largely due to the complexity and diversity of the high frequency spectra of individual sources. To understand the nature of the faint, high frequency  population and constrain the models better, a multifrequency study is required. In this paper we describe just such a study, examining the radio properties of a sub-sample of 10C sources in the Lockman Hole, a region which has been observed over a wide range of wavelengths.

The Lockman Hole is a region of the sky centred near 10$^{\rm h}$45$^{\rm m}$, +58$^{\circ}$ (J2000 coordinates, which are used throughout this work) with exceptionally low HI column density \citep{1986ApJ...302..432L}. The low infrared background (0.38 MJy~sr$^{-1}$ at 100~$\muup$m; \citealt{2003PASP..115..897L}) in this area of the sky  makes it ideal for infrared observations. As a result, as part of the Spitzer Wide-area Infrared Extragalactic survey (SWIRE; \citealt{2003PASP..115..897L}) sensitive infrared observations of $\approx$ 14 deg$^2$ of the Lockman Hole area have been made. The availability of deep infrared observations in the Lockman Hole has triggered deep observing campaigns at optical, X-ray and radio wavelengths.

The availability of data at such a wide range of frequencies makes the Lockman Hole a particularly good area for study. Here, the radio properties of sources detected in the 10C survey at 15.7~GHz in the Lockman Hole are investigated. The 10C data are combined with those from deep surveys at 610 MHz made with the Giant Meterwave Radio Telescope (GMRT; \citealt{2008MNRAS.387.1037G,2010BASI...38..103G}) and at 1.4~GHz made with the Westerbork Synthesis Radio Telescope (WSRT; \citealt{Guglielmino2012}), along with other available data over a range of radio frequencies. The surveys used are described in more detail in Section \ref{section:samples} along with the process for matching them to the 10C catalogue. Section \ref{section:source-parameters} describes how the parameters of the 10C sources are investigated using these surveys, including calculating the radio spectral indices and investigating the extent of the radio emission. The data are analysed and the properties of the 10C sources are discussed in Section \ref{section:analysis}. The 10C source population is compared to samples selected at 1.4~GHz in Section \ref{section:1.4-selected} and to the \citet{2008MNRAS.388.1335W,2010MNRAS.405..447W} simulated model of the radio sky in Section \ref{section:s3}. Optical identifications, redshift estimates and detailed discussions of the source types will be presented in a separate paper.

Throughout this paper the term `flat spectrum' refers to an object with spectral index $\alpha \leqslant 0.5$ and `steep spectrum' to an object with $\alpha > 0.5$.

\section{Sample selection}\label{section:samples}

\subsection{Surveys used}\label{section:rdataused}

\begin{figure}
\centerline{\includegraphics[bb=64 161 545 661,clip=,width=8.2cm]{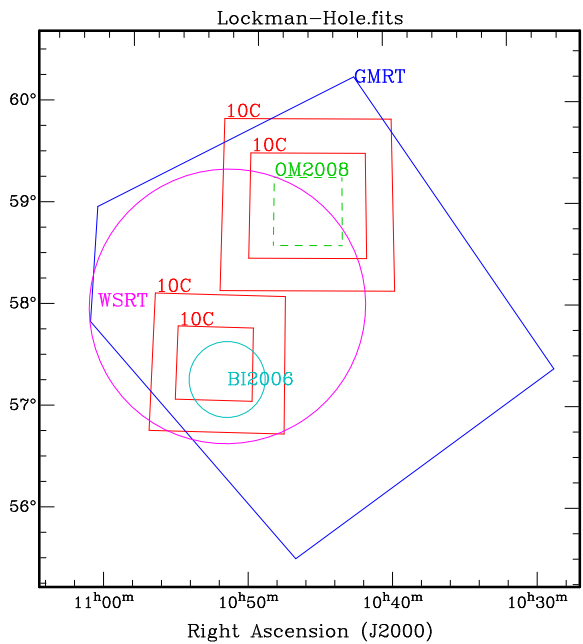}}
\caption{The deep radio surveys in the Lockman Hole region. The two larger (red) squares show the shallow region of the 10C survey fields, complete to 1~mJy. The two smaller squares contained within the larger squares are the deep 10C regions, complete to 0.5~mJy. The pentagon (dark blue) shows the GMRT survey area, the dashed square (green) shows the OM2008 survey area, the smaller circle (pale blue) shows the BI2006 survey area and the larger circle (pink) shows the WSRT survey area.  FIRST, NVSS and WENSS are not shown as they cover the whole region. See Table \ref{tab:surveys} for details of the different surveys shown.}\label{fig:source_map}
\end{figure}

The 10C radio survey at 15.7 GHz was made with the Arcminute Microkelvin Imager (AMI; \citealt{2008MNRAS.391.1545Z}) with a beam size of 30 arcsec. It covers $\approx$27 deg$^2$ complete to 1~mJy and $\approx$12 deg$^2$ complete to 0.5 mJy across ten different fields; a full description of the 10C survey and the source catalogue can be found in \citet{2011MNRAS.415.2699F} and \citet{2011MNRAS.415.2708D}. Two of the fields are in the Lockman Hole, covering an area of 4.64 deg$^2$ (see Fig. \ref{fig:source_map}) and detecting a total of 299 sources.

To investigate the properties of these 10C sources we have matched the 10C catalogue to other lower-frequency, but usually higher-resolution, radio catalogues as detailed below. This not only enables us to determine the radio spectral properties of the sources but also allows us to investigate the extent and structure of the radio sources in more detail. The greater positional accuracy of these higher-resolution catalogues is also vital for finding the counterparts of the 10C sources at optical and infrared wavelengths; the analysis of the optical and infrared properties of the 10C sources will be presented in a later paper.  The other radio surveys of the Lockman Hole used in this work are listed in Table \ref{tab:surveys} and briefly described below.

A series of deep observations at 610 MHz made with the GMRT covers the whole 10C Lockman Hole area except for a small corner of the field containing five 10C sources (Fig. \ref{fig:source_map}).  The GMRT image has an rms noise of $\approx$60 $\muup$Jy per beam in the central area (see \citealt{2008MNRAS.387.1037G,2010BASI...38..103G}, for details of the data reduction and source extraction). This deep image and the catalogue derived from it are used here.

A deep survey at 1.4~GHz carried out with the WSRT overlaps a large portion of the 10C survey (Fig. \ref{fig:source_map}) in the Lockman Hole \citep{Guglielmino2012}. The rms noise in the centre of the map is $\approx$11 $\muup$Jy; this map and the associated source catalogue are used in this study.

We also make use of several other catalogues and surveys covering the Lockman Hole region. The Faint Images of the Radio Sky at Twenty cm (FIRST; \citealt{1997ApJ...475..479W}), NRAO VLA Sky Survey (NVSS; \citealt{1998AJ....115.1693C}) and Westerbork Northern Sky Survey (WENSS; \citealt{1997A+AS..124..259R}) cover the whole area. There are also several deep observations made with the VLA within the Lockman Hole area: \citet{2009AJ....137.4846O} at 324 MHz (OMK2009) and \citet{2006MNRAS.371..963B} and \citet{2008AJ....136.1889O} at 1.4 GHz (BI2006 and OM2008). The locations of these surveys are shown in Fig. \ref{fig:source_map} and are summarised in Table \ref{tab:surveys}.

\begin{figure}
\centerline{\includegraphics[bb=62 184 567 662,clip=,width=8.2cm]{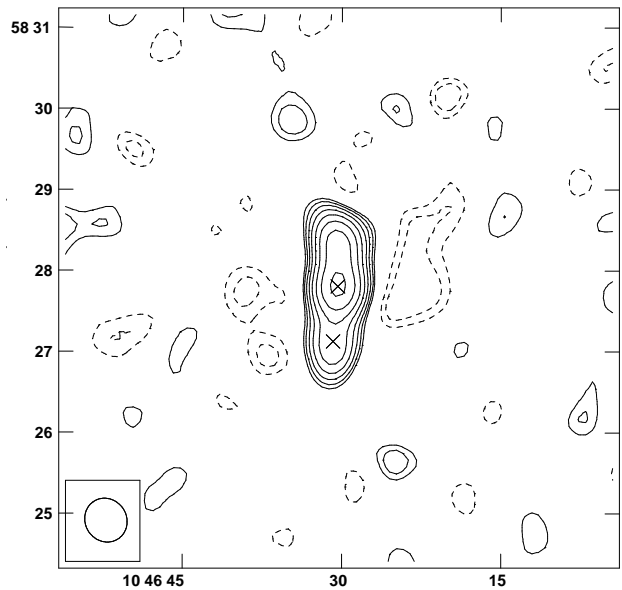}}
\caption{15.7-GHz image of an example extended source listed as two components in the 10C catalogue. The contours are plotted at ($\pm 2\sqrt{2^n}, n = 0, 1... 7) \times 0.136 \textrm{ mJy}$. The crosses mark the positions of the two sources in the 10C catalogue. There are three such pairs of sources in the 10C Lockman Hole fields; in each case the flux densities of the separate components are combined.}\label{fig:10C-extended}
\end{figure}

Sixteen 10C sources in the Lockman Hole are labelled in the 10C catalogue as part of a `group' (see \citealt{2011MNRAS.415.2708D} for details) -- in this case consisting of eight pairs. Contour maps of these eight pairs were examined by eye, along with images of their counterparts in FIRST and/or GMRT. For three of these pairs, there is evidence of structure connecting the two components, so it was decided to combine the two components into one source (an example of one such source is shown in Fig. \ref{fig:10C-extended}).The positions listed in the catalogue for these three pairs is the point mid-way between the two components. For the remaining five pairs the components were left as separate sources. This leaves a total of 296 sources in the 10C sample.

These 296 sources, selected at 15.7 GHz, are the subject of this paper. A subsample of 89 sources, with flux densities greater than 0.5~mJy, selected from the region complete to 0.5~mJy is defined as `Sample A'. A second subsample of 118 sources, `Sample B', which has some overlap with Sample A, forms a sample complete to 1~mJy. Sixty-two 10C sources, 43 of which form a sample complete to 0.5~mJy at 15.7~GHz, are in the deep areas observed with the VLA at 1.4~GHz by \citeauthor{2006MNRAS.371..963B} and \citeauthor{2008AJ....136.1889O} (BI2006 and OM2008). All 62 sources are detected at 1.4~GHz so spectral information is available for all of them. They therefore form a particularly useful subsample, defined as Sample C. Table \ref{tab:samples} contains a summary of the different subsamples used in this paper.

\subsection{Matching the radio catalogues}\label{section:matching}

\subsubsection{Choosing a match radius}\label{section:match-radius}

\begin{figure}

\centerline{\includegraphics[width=8.2cm]{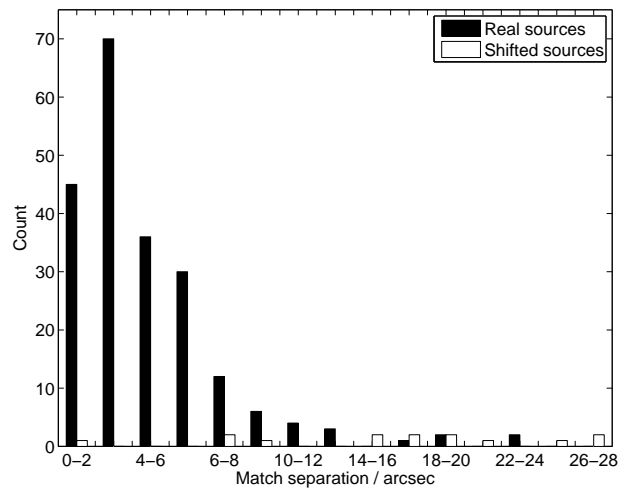}}

\caption{The number of matches with a given separation between the 10C and GMRT catalogues for the true and shifted distribution of 10C sources. Beyond 15 arcsec the number of true and shifted matches becomes comparable so 15 arcsec is chosen as the match radius.}\label{fig:radiomatches}

\end{figure}

The Topcat software package\footnote{see: http://www.starlink.ac.uk/topcat/} was used to match the catalogues. The match radius was chosen so as to maximise the number of real associations and avoid false matches. The error in the 10C positions is $\approx$~6~arcsec, which is larger than the errors in the other catalogues used here (for example the error in the GMRT positions is $\lesssim$~1~arcsec), so the errors in the 10C positions tend to dominate. Here we describe the process by which we chose a suitable match radius and assessed the probability of genuine and random matches for the 10C and GMRT catalogues. To create a random distribution of sources the 10C sources were shifted by 5 arcmin in declination. Both this shifted catalogue and the true 10C catalogue were then matched to the GMRT catalogue and the angular separation between a shifted or true 10C source and its nearest GMRT source was recorded. Once a shifted or true 10C source was matched to a source in the GMRT catalogue no further matches were sought for that source. The matches were then binned according to their separation for both the true and shifted sources. The resulting distribution is shown in Fig. \ref{fig:radiomatches}. It can be seen that beyond 15 arcsec the numbers of random matches become comparable for both the true and shifted catalogues; thus 15 arcsec was chosen as the match radius. The number of matches with the shifted distribution within a separation of 15 arcsec is 4 (out of 296 sources), meaning the probability of a random match within 15 arcsec is $\approx$1.5 percent of the total number of sources being matched. 15 arcsec corresponds to 2.5$\sigma$, where $\sigma$ is the typical error in the 10C source positions, suggesting that the matching is limited by the error in the 10C source positions.

This process was repeated for the other catalogues and it was decided that 15 arcsec was a suitable match radius to use when matching each of the catalogues to the 10C catalogue. The number of random matches within 15 arcsec for each catalogue is shown in Table \ref{tab:matching}.

\begin{table*}
\caption{Radio catalogues in the Lockman Hole}\label{tab:surveys}
\bigskip
\begin{tabular}{lllddld}\hline
 Catalogue & Reference(s) & Epoch of observation & \dhead{Frequency} & \dhead{Beam size} & rms noise & \dhead{Area covered}\\
 &  &  & \dhead{/GHz} & \dhead{/arcsec} & /mJy & \dhead{/deg$^2$}\\\hline
  10C -- shallow  & \twolines{\citet{2011MNRAS.415.2699F}}{\citet{2011MNRAS.415.2708D}} & Aug 2008 -- June 2010  & 15.7 & 30 & 0.1 & \dhead{4.64 (deep areas included)} \\
  10C -- deep     & \twolines{\citet{2011MNRAS.415.2699F}}{\citet{2011MNRAS.415.2708D}} & Aug 2008 -- June 2010  & 15.7 & 30 & 0.05 &  1.73 \\
  GMRT    & \twolines{\citet{2008MNRAS.387.1037G}}{\citet{2010BASI...38..103G}} & Jul 2004 -- Oct 2006 & 0.610 & \dhead{$6\times5$} & 0.06 & 13 \\
  WSRT    & \citet{Guglielmino2012}                     & Dec 2006 -- Jun 2007   & 1.4   & \dhead{$11\times9$}  & 0.011  & 6.6  \\
  OM2008  & \citet{2008AJ....136.1889O}                 & Dec 2001 -- Jan 2004   & 1.4   & 1.6 & 0.0027 & 0.011  \\
  OMK2009 & \citet{2009AJ....137.4846O}                 & Feb 2006 -- Jan 2007   & 0.324 & 6   & 0.07   & 3.14 \\
  BI2006  & \citet{2006MNRAS.371..963B}                 & Jan 2001 -- Mar 2002   & 1.4   & 1.3 & 0.0046 & 0.089  \\
  FIRST   & \citet{1997ApJ...475..479W}                 & 1997 -- 2002   & 1.4   & 5   & 0.15   & \dhead{Whole area}   \\
  NVSS    & \citet{1998AJ....115.1693C}                 & 1997  & 1.4   & 45  & 0.45   & \dhead{Whole area}   \\
  WENSS   & \citet{1997A+AS..124..259R}                 & 1991 -- 1996   & 0.325 & 54  & 3.6    & \dhead{Whole area}   \\\hline
\end{tabular}
\end{table*}

\begin{table*}
\caption{The different subsamples in the Lockman Hole used in this paper}\label{tab:samples}
\medskip
\begin{center}
\begin{tabular}{p{2cm}p{10cm}d}\hline
Sample      & Description & \dhead{Number of sources}\\\hline
All sources & All sources detected at 15.7~GHz. Includes sources below 0.5~mJy. & 296\\
Sample A    & Complete to 0.5~mJy at 15.7~GHz & 89\\
Sample B    & Complete to 1.0~mJy at 15.7~GHz & 118\\
Sample C    & All sources in the deep regions surveyed at 1.4~GHz with the VLA by \citet{2008AJ....136.1889O} and \citet{2006MNRAS.371..963B}. This sample contains sources below the completeness limit at 15.7~GHz. & 62\\\hline
\end{tabular} 
\end{center}
\end{table*}

\subsubsection{Results of the matching}\label{section:matching-results}

A summary of the matching between the 10C sources and the other radio catalogues is shown in Table \ref{tab:matching}. In total, 266 out of the 296 10C sources are detected at at least one other wavelength. Table \ref{tab:multi-matches} summarises the number of sources with matches in more than one other catalogue. 

\begin{table}
\caption{A summary of the matching between the 10C catalogue and other radio catalogues in the Lockman Hole. Column (2) shows the number of matches to 10C sources within 15 arcsec and column (3) shows the number of 10C sources in the survey area. Column (4) shows the number of random matches within 15 arcsec, i.e. the number of matches to the shifted 10C sources.}\label{tab:matching}
\medskip
\begin{center}
\begin{tabular}{p{2cm}ddd}\hline
  Catalogue & \dhead{Number of} & \dhead{Number of}   & \dhead{Number of} \\
                    & \dhead{matches}   & \dhead{10C sources} & \dhead{random matches}\\
   (1)  & (2) & (3) & (4)\\\hline
GMRT    & 205 & 291 & 8\\
WSRT    & 160 & 182 & 10\\
FIRST   & 196 & 296 & 0\\
NVSS    & 166 & 296 & 0\\
WENSS   & 86  & 296 & 6\\
OM2008  & 27  & 27  & 2\\
OMK2009 & 116 & 156 & 2\\
BI2006  & 35  & 35  & 3\\\hline
\end{tabular}
\end{center}
\end{table}

\begin{table}
\caption{A summary of the number of 10C sources with multi-frequency matches.}\label{tab:multi-matches}
\medskip
\begin{center}
\begin{tabular}{p{2.5cm}d}\hline
Number of matches & \dhead{Number of sources}\\\hline
10C only       & 30\\
One match only & 46\\
Two matches    & 36\\
Three matches  & 27\\
Four matches   & 55\\
Five matches   & 78\\
Six matches    & 23\\
Seven matches  & 1\\\hline
\end{tabular}
\end{center}
\end{table}

\begin{figure}
\centerline{\includegraphics[width=8.2cm]{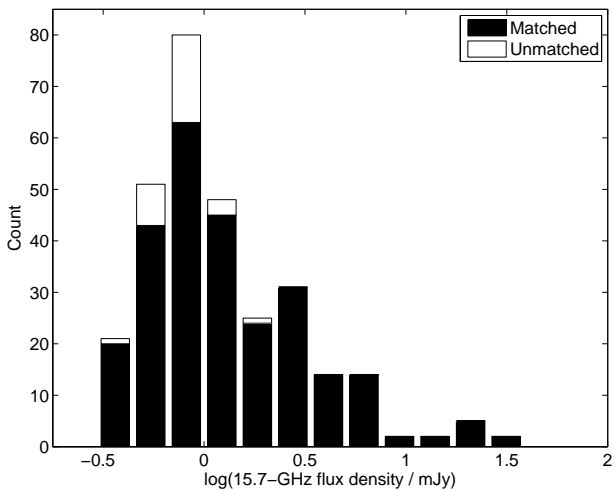}}
\caption{The distribution of 10C sources with and without associations in the other radio catalogues as a function of flux density. This diagram contains all 296 sources detected, some of which are below the completeness levels of the 10C survey. All of the 10C sources with no matches have a flux density below 2 mJy.}\label{fig:10Cflux}
\end{figure}

 For the 30 10C sources with no other detections, upper limits on the flux  density at 610~MHz from GMRT and at 1.4~GHz from WSRT and FIRST can be used, as described in Section \ref{section:flux-values}. All of the 10C sources with no other detections have a low flux density (see Fig. \ref{fig:10Cflux}), with the majority below 1~mJy and all below 2~mJy.

\section{Deriving the source parameters}\label{section:source-parameters}

\subsection{Flux density values}\label{section:flux-values}

\subsubsection{GMRT}\label{section:res_GMRT}

\begin{figure*}
\centerline{\includegraphics[bb=48 177 567 695,clip=,width=5.5cm]{10CJ104320+585621.PS} \quad
            \includegraphics[bb=40 192 567 715,clip=,width=5.5cm]{match-045.eps} \quad
            \includegraphics[bb=40 192 567 715,clip=,width=5.5cm]{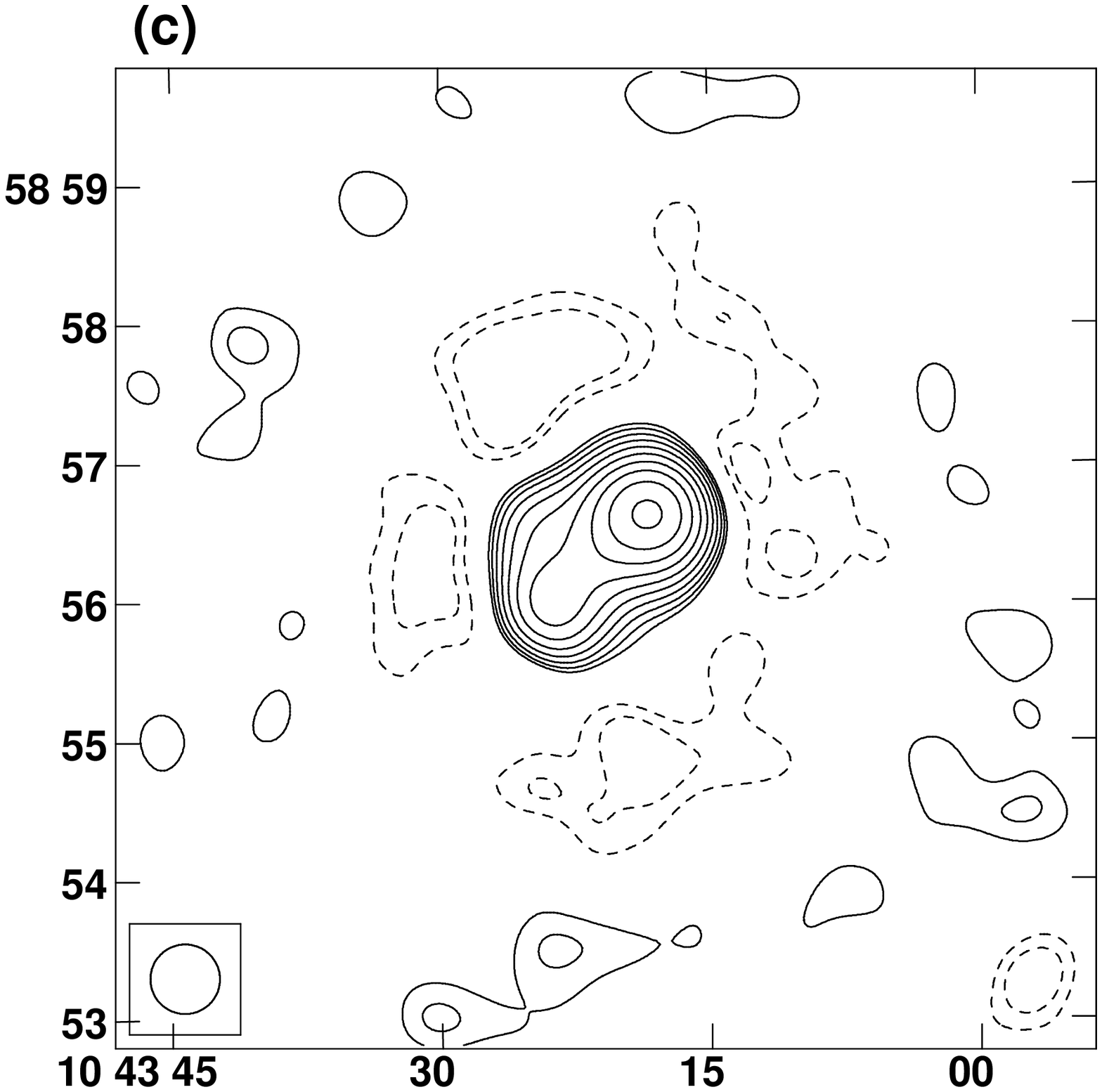}}

\bigskip

\centerline{\includegraphics[bb=48 177 567 695,clip=,width=5.5cm]{10CJ104336+592618.PS} \quad
            \includegraphics[bb=40 192 567 715,clip=,width=5.5cm]{match-048.eps} \quad
            \includegraphics[bb=40 192 567 715,clip=,width=5.5cm]{smoo-048.eps}}
\caption{The difference in resolution between AMI Large Array (used for the 10C survey) and the GMRT. Figs. (a) and (d) each show a source imaged with AMI, Figs. (b) and (e) show the same sources imaged with GMRT. In both cases several sources in the GMRT catalogue correspond to one source in the 10C catalogue, so care must be taken when comparing the catalogues. In Figs. (c) and (f) the GMRT images have been smoothed to create an image with a resolution comparable to the 10C images. This allows for a more direct comparison between the two surveys. The contours are drawn at $(\pm 2\sqrt{2^n}, n = 0, 1... 7) \times x \textrm{ mJy}$ where $x$ = 0.074 for (a), 0.1 for (b), 0.3 for (c), 0.086 for (d), 0.16 for (e) and 0.3 for (f). \label{fig:resolution}}
\end{figure*}

The large difference in resolution between the 10C (beam size 30 arcsec) and GMRT (beam size 6 arcsec) observations means that care must be taken when comparing flux densities. As illustrated in Fig. \ref{fig:resolution}, many 10C sources are extended or resolved into multiple components at 610 MHz. The components of such sources will be listed as separate entries in the GMRT catalogue. In addition, for some of the sources in the GMRT catalogue there may be extended low brightness structure too faint to be seen in the high resolution images. In order to reduce these problems, contour plots of all the GMRT sources which matched to a source in the 10C catalogue (using a 15 arcsec match radius as described in Section \ref{section:match-radius}) were examined by eye. For the 50 sources which appeared multiple or extended, the GMRT images were convolved in AIPS\footnote{Astronomical Image Processing System, see http://www.aips.nrao.edu/.} to a 30-arcsec gaussian to create an image of comparable resolution to the 10C data (see Fig. \ref{fig:resolution}). The total 610-MHz flux density of each source (which had already been matched to a 10C source from the higher resolution image) was then estimated from the smoothed GMRT image. The integrated 610-MHz flux densities were estimated by fitting a Gaussian using the AIPS task JMFIT. In the few cases where JMFIT did not converge (due to the presence of another bright source in the subimage) the integrated flux density was found by hand using the AIPS task TVSTAT.

For the 84 10C sources for which a counterpart was not present in the full resolution GMRT catalogue the GMRT images were used to place an upper limit on the source flux density at 610~MHz. A sub image of 2.5~arcmin was extracted and smoothed (as described above) to check for any large scale structure which might have been resolved out in the original image. In eight cases a source could be seen in the smoothed GMRT images with a peak within 15 arcsec of the 10C source position; for these sources a value for the integrated flux density was obtained manually using TVSTAT. For the remaining sources, the upper limit on the flux density was taken to be three times the noise in this smoothed image. It was not possible to get any information about the flux density at 610 MHz for the five 10C sources outside the GMRT area so these sources are excluded from all discussions relating to 610-MHz data.

\subsubsection{NVSS}\label{section:NVSS}

NVSS has a resolution comparable to the 10C survey so the flux densities can be directly compared without any problems caused by resolution differences. Because of this, it is useful to be able to calcuate spectral indices using NVSS and 10C values only, so upper limits were found for all sources not matched in the NVSS catalogue despite the fact that some of these sources have a counterpart in one of the deeper 1.4-GHz surveys used. NVSS images of all 10C sources without a match in the NVSS catalogue were examined to see if there was a source present which was below the catalogue limit and for the few sources where this was the case, the flux density of the NVSS source was found manually using the AIPS task TVSTAT. For the remaining unmatched sources, an upper limit of three times the local noise in the NVSS image was placed on the flux density.

\subsubsection{WSRT and FIRST}\label{section:WSRT-FIRST}

As the WSRT synthesised beam (12 arcsec) is smaller than the 10C synthesised beam (30 arcsec), several sources were resolved into multiple components in the WSRT catalogue but appeared as a single component in the 10C catalogue. In order to reduce the effects of resolution when comparing the flux densities, sub-images of the WSRT image at the position of each 10C source were examined by eye; for those sources where there were multiple WSRT components associated with one 10C source, the flux densities of the WSRT components were added together. (In fact, in all cases except one, the WSRT sources in question were listed as a multi-component source in the WSRT catalogue and the total flux  density from the catalogue was used). This process was also carried out for FIRST sources.

For those 10C sources in the WSRT survey area without a match to the WSRT catalogue, the WSRT image was examined to see if there was a source present which was below the catalogue limit. This was not the case for any of the unmatched 10C sources in the WSRT area. The upper limit of flux density was taken to be three times the local noise in the WSRT image. For the 28 sources which were unmatched at 1.4 GHz and outside the WSRT area, an upper limit of 1~mJy was placed on the flux density because this is the FIRST completeness limit and these sources are all within the FIRST survey area but not detected.

\subsubsection{OM2008 and BI2006}

The two deep 1.4~GHz surveys made with the VLA (OM2008 and BI2006) have very small synthesised beams ($\approx 1.5$ arcsec) compared to the 10C survey. As images are not available for these surveys, the flux density values from the catalogues were used. OM2008 do account for the fact that some of the sources may be extended when calculating flux densities. They convolve the full resolution images to an effective beam size of 3, 6 and 12 arcsec and compare the flux densities derived from the four images. BI2006 do not attempt to account for extended sources when calculating the flux densities, however there is only one 10C sources which has a counterpart in BI2006 and not in any other 1.4-GHz catalogue so this will not significantly affect these results.

\subsection{The effects of variability}\label{section:variability}

The different surveys used in this paper were not carried out simultaneously so it is important to consider the possible effects of variability on the observed spectral index distributions.  The epoch of observation for each survey is shown in Table \ref{tab:surveys}.  The time interval between the 15-GHz and the 610-MHz observations is in the region of 4 -- 6 years and between the 15-GHz and 1.4-GHz observations 5 -- 10 years.

Whilst there is currently no data on the variability of sources at 15~GHz at the low flux density end of our sample, there have been some systematic studies at higher flux densities. \citet{2006MNRAS.370.1556B} studied the 15-GHz variability of 51 9C sources with flux densities $> 25$ mJy over a 3-year period.  They found that while there was no evidence for variability (above the $\sim$6 percent flux calibration uncertainties) in steep spectrum sources, half of the flat spectrum objects were variable. In total, 29 percent of the sources studied were found to vary. \citet{2006MNRAS.371..898S} observed 173 20-GHz sources with flux densities $> 100$ mJy over a 2-year period and found that 42 percent varied by  more than 10 percent. However, they found no correlation between variability and radio spectral index. More recently, \citet{2011MNRAS.416..559B} investigated the variability of 159 sources with 20-GHz flux densities $> 200$ mJy and found the variability to be slightly larger than that found by \citeauthor{2006MNRAS.371..898S}, with an r.m.s. amplitude of 38 percent at 20~GHz on a timescale of a few years. \citeauthor{2011MNRAS.416..559B} note that there is some indication that the variability decreases as flux density decreases. \citet{2011MNRAS.415.1597M} studied the variability of a brighter sample ($S_{20 \rm{GHz}} > 500$~mJy) at 20~GHz and found similar levels of variability to \citeauthor{2006MNRAS.371..898S}  At higher flux densities still \citet{2009MNRAS.400..995F} have looked at variability at 16~GHz in a complete sample of 97 sources with flux densities $> 1$~Jy over timescales of about 1.5~years.  They found that 15 percent of the sources vary by more than 20 percent; however, in contrast to the results of \citeauthor{2006MNRAS.371..898S} but in agreement with those of \citeauthor{2006MNRAS.370.1556B}, the spectra of the variable sources are flatter than those of the non-variable ones.

The variability properties of the population studied here are not known. However, on the assumption that the faint sources in the 10C sample exhibit the same sort of flux densiy variations as shown in the higher flux density samples discussed above, a significant fraction of them are likely to have varied over the period between the observations. Thus the spectral indices of individual objects may be unreliable. However, given that the sources are probably equally likely to increase or decrease in flux density this should not have a major effect on the overall spectral index distribution.

\subsection{Spectral indices}\label{section:alpha-methods}

To investigate the spectral properties of the source sample in a quantitative way, the spectral index was calculated between 15.7~GHz and 1.4~GHz ($\alpha_{1.4}^{15.7}$) and 15.7~GHz and 610~MHz ($\alpha_{0.61}^{15.7}$) for each source. For $\alpha_{1.4}^{15.7}$ all 296 sources are studied, for $\alpha_{0.61}^{15.7}$ the 5 sources outside the GMRT area are excluded as there is no 610-MHz flux density information available.  For the sources with no match in GMRT, a limiting spectral index was calculated from the upper limit placed on the flux density from the GMRT image, as described in Section \ref{section:res_GMRT}. 

The distributions of $\alpha_{1.4}^{15.7}$ were investigated in two ways, using slightly different procedures, to check the effects of resolution on the data. The first, $\alpha_{1.4}^{15.7}M$,  makes use of all of the 1.4-GHz data available; for the sources where there is more than one 1.4-GHz flux density, flux densities are chosen according to resolution in the following order of preference: NVSS, WSRT, OM2008, FIRST, BI2006; FIRST and BI2006 are last as they are the most likely to resolve out some of the flux of the 10C sources because of their small beam sizes (5 and 1.5~arcsec respectively). OM2008 also has a small beam (1.6 arcsec) but the sources have been convolved with gaussians of varying radius to try and overcome the resolution problem. NVSS and WSRT have larger beam sizes (30 and 12 arcsec respectively) which are more comparable to the 10C beam. For the sources with no match in any of the 1.4-GHz catalogues which are in the WSRT survey area, the upper limit from the WSRT image (as described in Section \ref{section:WSRT-FIRST}), is used to calculate a limiting spectral index. For the remaining sources, an upper limit of 1 mJy from the FIRST survey is used. 

The second value, $\alpha_{1.4}^{15.7}N$,  only uses values from the NVSS catalogue as this has a resolution comparable to the 10C survey. For the 10C sources which do not appear in NVSS, the limit is derived from the NVSS image (see section \ref{section:NVSS}). These values of the spectral index contain a larger number of upper limits but provide a useful comparison when considering the effects of resolution.

 Possible effects of variability on the spectral indices are discussed in Section \ref{section:variability}.

\subsection{Extent of the radio emission}\label{section:extended-methods}

\begin{figure}
\centerline{\includegraphics[width=8.2cm]{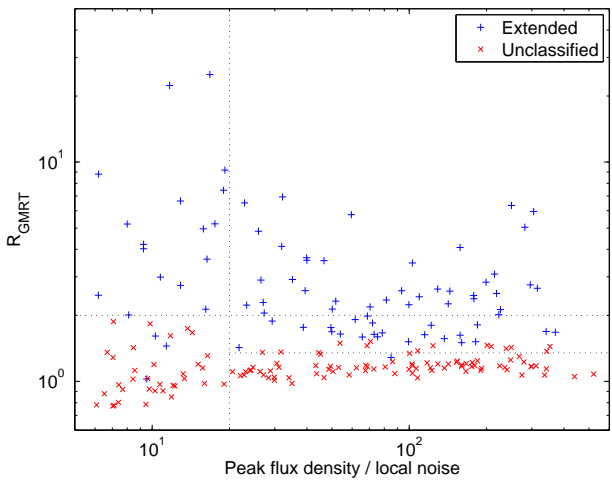}}

\caption{Compactness ratio calculated using GMRT data, $R_{\rm GMRT}$, against the signal to noise ratio from the GMRT catalogue. The vertical line shows the divide at peak flux  density / noise = 20 and the horizontal lines at $R = 1.35$ and 2.00 indicate the cutoffs in $R$ used when classifying the extended sources, see text for details. \label{fig:R_GMRT}}
\end{figure}

\begin{figure}
\centerline{\includegraphics[width=8.2cm]{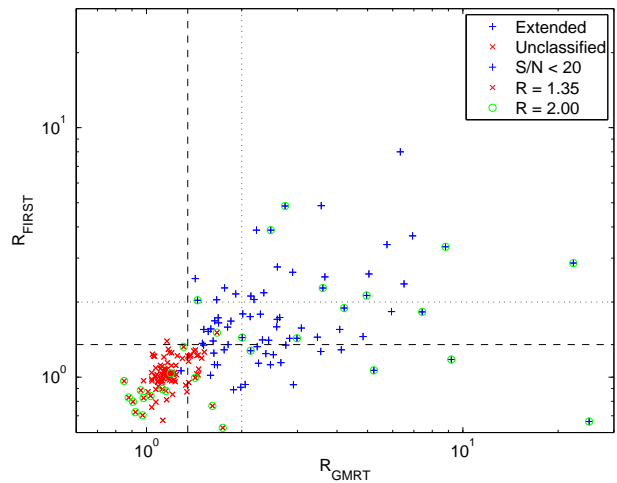}}

\caption{Value of the compactness ratio, $R$, calculated using FIRST and GMRT data. The horizontal and vertical lines indicate the values used when identifying the extended sources; see text for details. Sources classified as extended and those not classified are shown separately. Sources with a signal to noise ratio $< 20$ in the GMRT data are indicated with a green circle.\label{fig:R}}
\end{figure}

To determine in a quantitative manner whether a source is extended or not, a compactness ratio $R$ is often used. This is usually taken as the ratio of the integrated or total flux density of a source $S_{int}$ and its peak flux density on a map $S_{peak}$ i.e. $R = S_{int}/S_{peak}$. For the 10C sample considered here we do not use the 10C observations to find $R$ because the large beam size of $\approx 30$ arcsec means that the majority of the sources are unresolved. Instead, we use the matched data from the lower frequency catalogues, which have smaller beam sizes and can therefore provide more information about the angular size of a source. Four values of $R$ are calculated, using data from the GMRT, FIRST and OMK2009 (324~MHz VLA observations) catalogues, which all have beam sizes $\approx 6$ arcsec, and the WSRT catalogue which has a beam size of 12 arcsec. The deep 1.4~GHz VLA observations (OM2008 and BI2006) are not used in this analysis because their beam sizes are considerably smaller ($\approx 1$ arcsec) so that the data are not comparable with those from the other catalogues used here.

To take account of the effects of noise, several other studies (e.g \citealt{2007A&A...463..519B}, \citealt{2006A&A...457..517P}) plot $R$ against the signal to noise ratio and fit a lower envelope to the data; reflecting this about $R=1$ gives a curve which provides the cutoff between the extended and compact sources. We have not used this method for the sample of sources in this paper due both to the relatively small number of sources in the sample and to the range of different surveys used. Instead, each source was examined by eye and the criteria described below were decided upon to identify the extended sources. 
\begin{figure}

\centerline{\includegraphics[width=8cm]{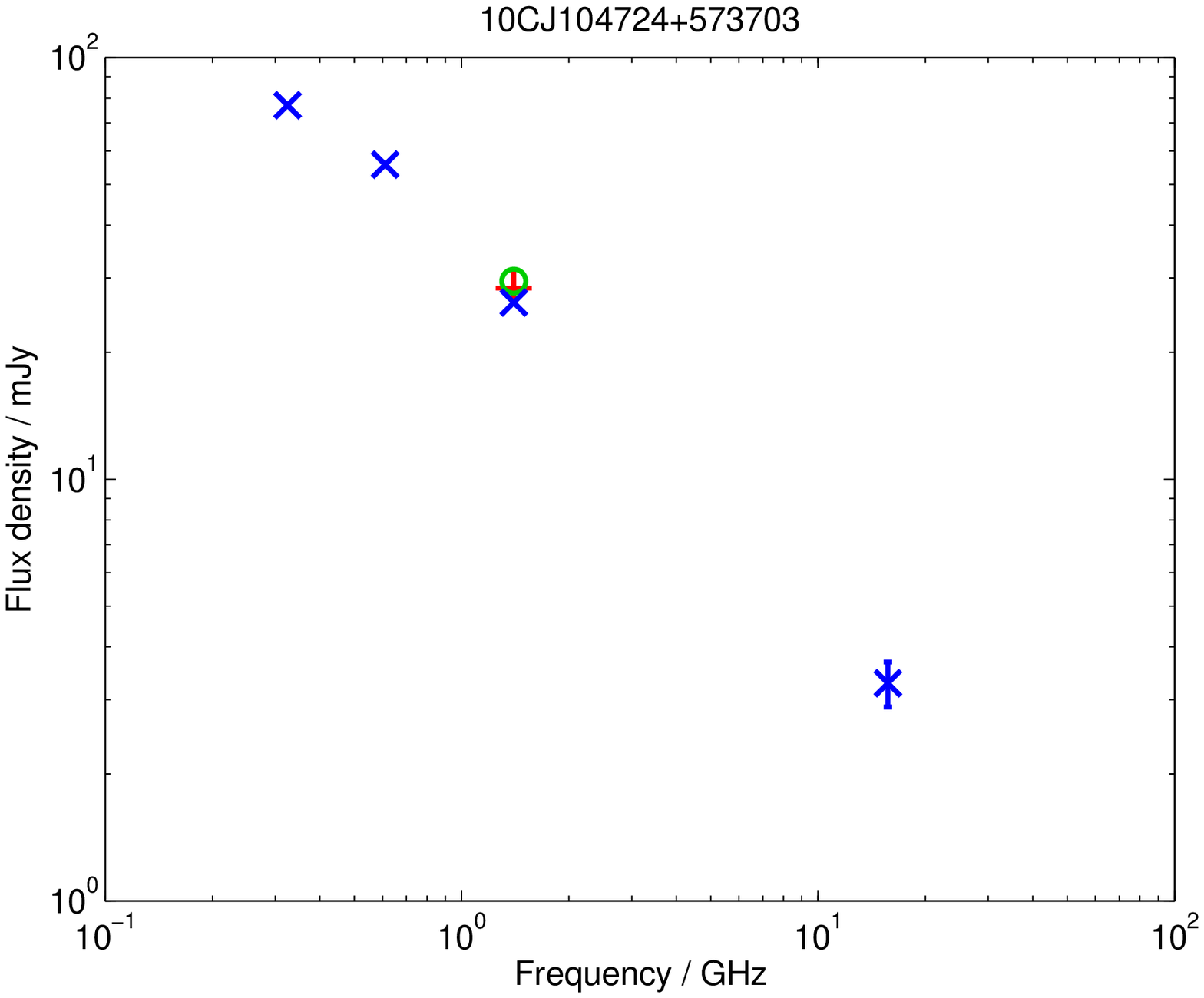}}
\medskip

\centerline{\includegraphics[width=8cm]{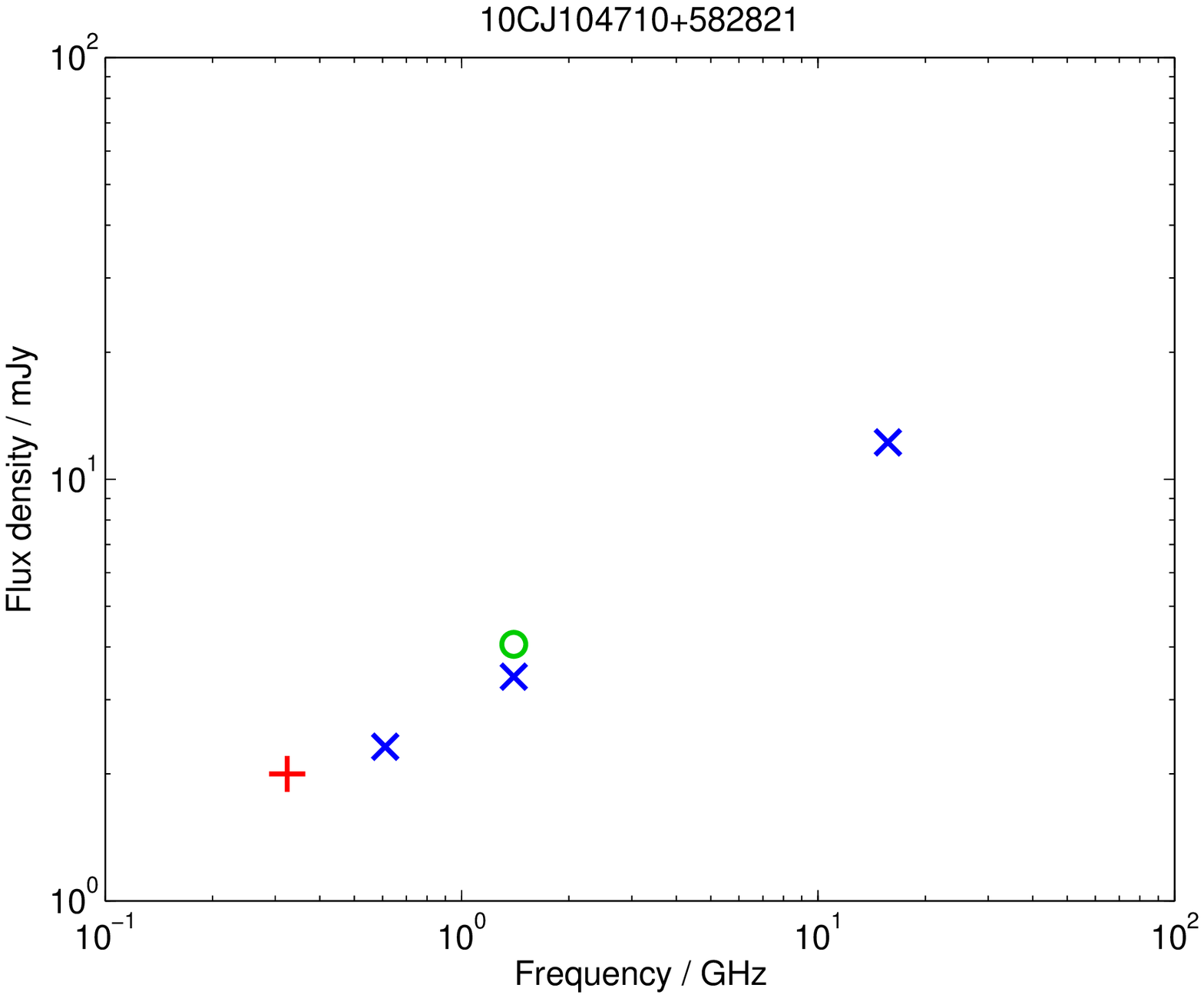}}
\medskip

\centerline{\includegraphics[width=8cm]{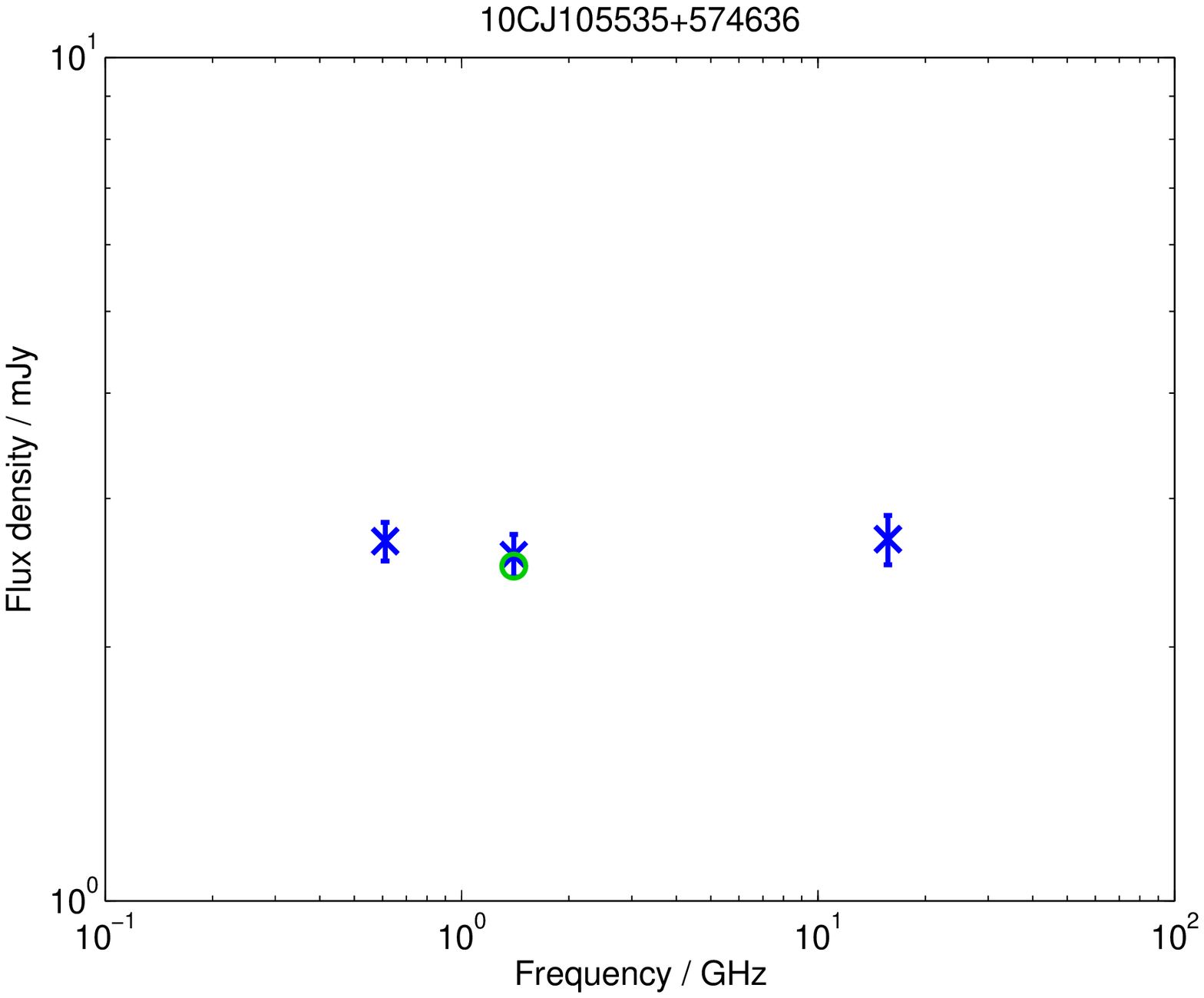}}
\caption{Example spectra. 10CJ104724+573703 is an example of a steep spectrum object, 10CJ104710+582821 is a rising spectrum object and  10CJ105535+574636 is an example of a flat spectrum object. The blue cross at 324~MHz (if present) is from WENSS, the red plus at 324~MHz is from OMK2009 and the value at 610~MHz is from GMRT. The blue cross at 1.4~GHz is from FIRST, the green circle is from WSRT and the red plus is from NVSS. The value at 15.7~GHz is from 10C. Error bars are not shown when they are smaller than the symbol plotted.\label{fig:spectra}}
\end{figure}

At lower signal to noise ratios the errors in the $R$ values become larger, due to the larger errors in the integrated and peak flux density values. This is evident in Fig. \ref{fig:R_GMRT} for the GMRT data, where the number of sources with values of $R < 1$, which must be due to errors in the flux density values, increases at lower signal to noise ratios. Therefore different criteria are used to identify the extended sources at high and low signal to noise. Sources were classified as extended if they had a signal to noise ratio greater than 20 in the GMRT data and values of $R > 1.35$ in at least two of the datasets used, or a value of $R > 2$ in at least one of the datasets. Sources with a signal to noise ratio less than 20 in the GMRT data were classified as extended if the value of $R$ was greater than 2 in any of the datasets. A comparison of two of the $R$ values used is shown in Fig. \ref{fig:R}. For most sources the two values are similar and some of the largest variations are for those sources with low signal to noise values, which is expected because the errors in the integrated flux densities are larger for the fainter sources.

We are confident that all the sources which fulfil these criteria are extended; however, as we have erred on the side of caution, there will be some sources which have not been classified as extended but do in fact have extended emission on the scale investigated here. In particular, sources which only have a value of $R$ from the WSRT catalogue are not classified as extended as the WSRT beam is larger than that of the other catalogues used here. For this reason, those sources which are not classified as extended are placed in the `unclassified' bin. The 36 sources which do not have a counterpart in any of the four catalogues used here are classified as `no information'. The criteria used to select the extended sources here is roughly equivelent to selecting sources with angular sizes larger than $\approx 6$ arcsec. 

A catalogue containing the flux density, spectral index and $R$ values will be available online.

\section{Sample analysis}\label{section:analysis}

\subsection{Radio spectral properties}\label{section:spectral}

\begin{figure}
\centerline{
\includegraphics[width=8.2cm]{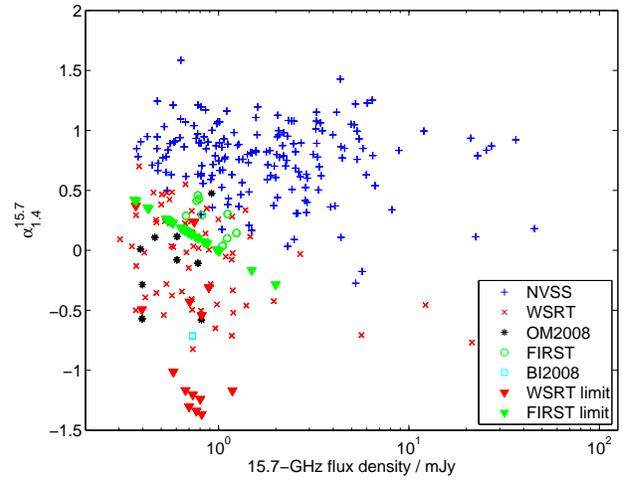}}
\caption{The spectral index $\alpha^{15.7}_{1.4}M$ against 15.7-GHz flux density, showing which catalogue the 1.4-GHz flux density values come from.  \label{fig:alpha-S}}

\end{figure}

\begin{figure*}
\centerline{
\includegraphics[width=8cm]{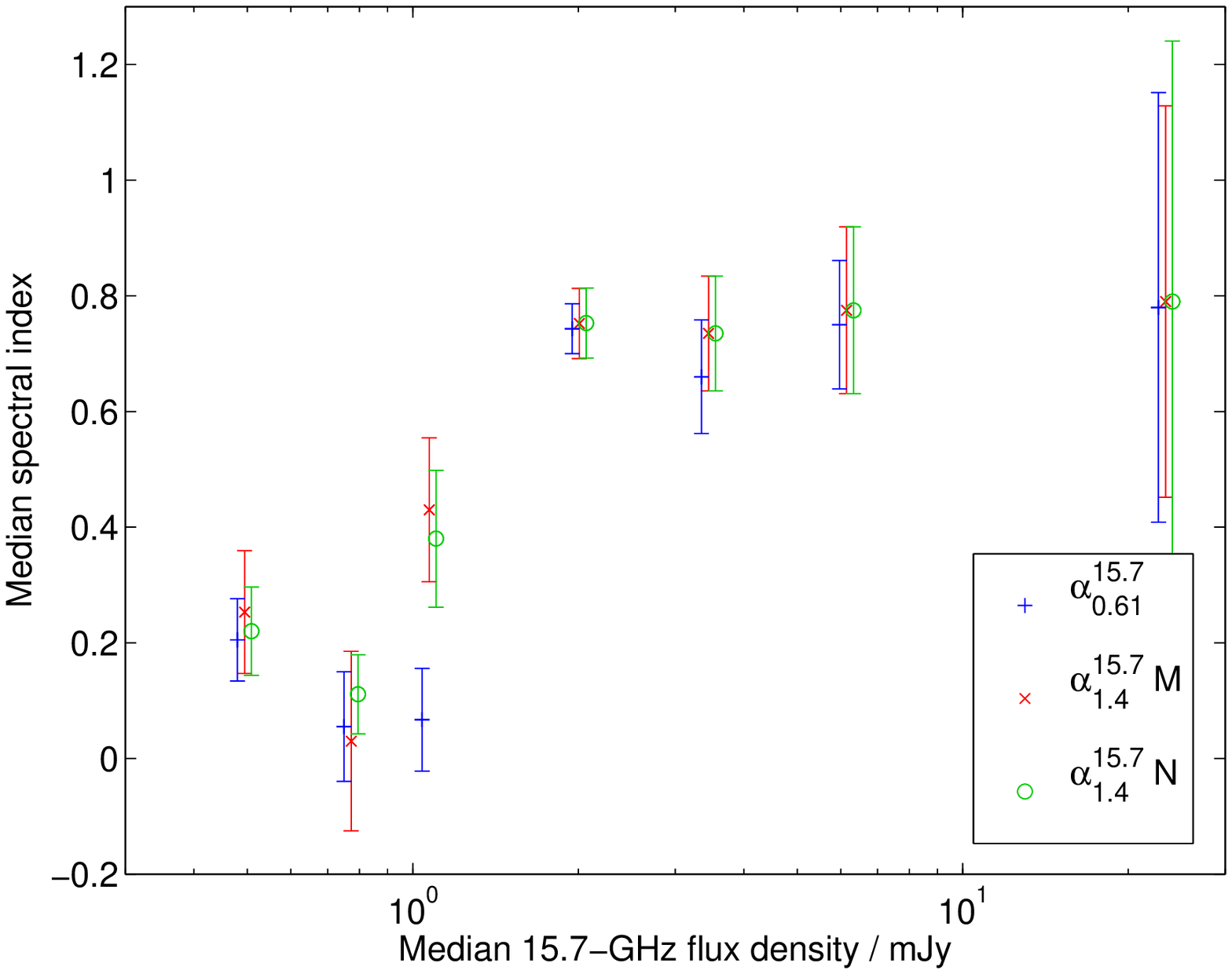}
\qquad
\includegraphics[width=8cm]{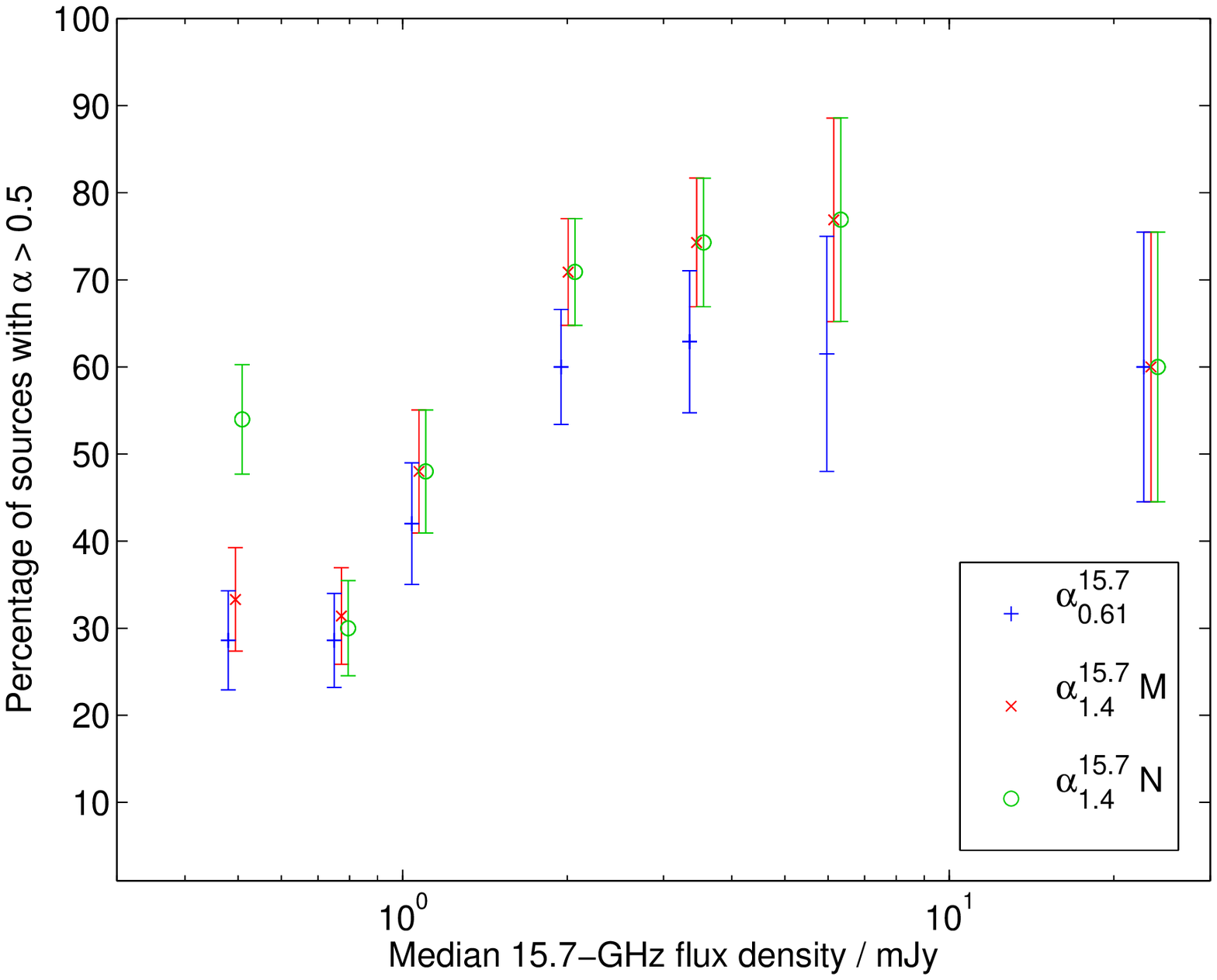}}
\caption{The median spectral index and percentage of sources with $\alpha > 0.5$ as a function of 15.7-GHz flux density for all 10C sources.  The values of $\alpha^{15.7}_{1.4}M$ are calculated using the best 1.4-GHz flux density values available (order of preference = NVSS, WSRT, OM2008, FIRST, BI2006) while the values of $\alpha^{15.7}_{1.4}N$ are calculated using 1.4-GHz flux density values from NVSS only. The points are plotted at the median flux density in each bin but are offset for clarity. Limiting spectral indices are included in the median using survival analysis (see text). The error bars in the median spectral index plot show the interquartile range divided by $\sqrt{N}$ to give an indication of the errors in the medians.  Note the large change in spectral index between 1 and 2 mJy. \label{fig:median-all}}

\end{figure*}

\begin{figure*}
\centerline{
\includegraphics[width=8cm]{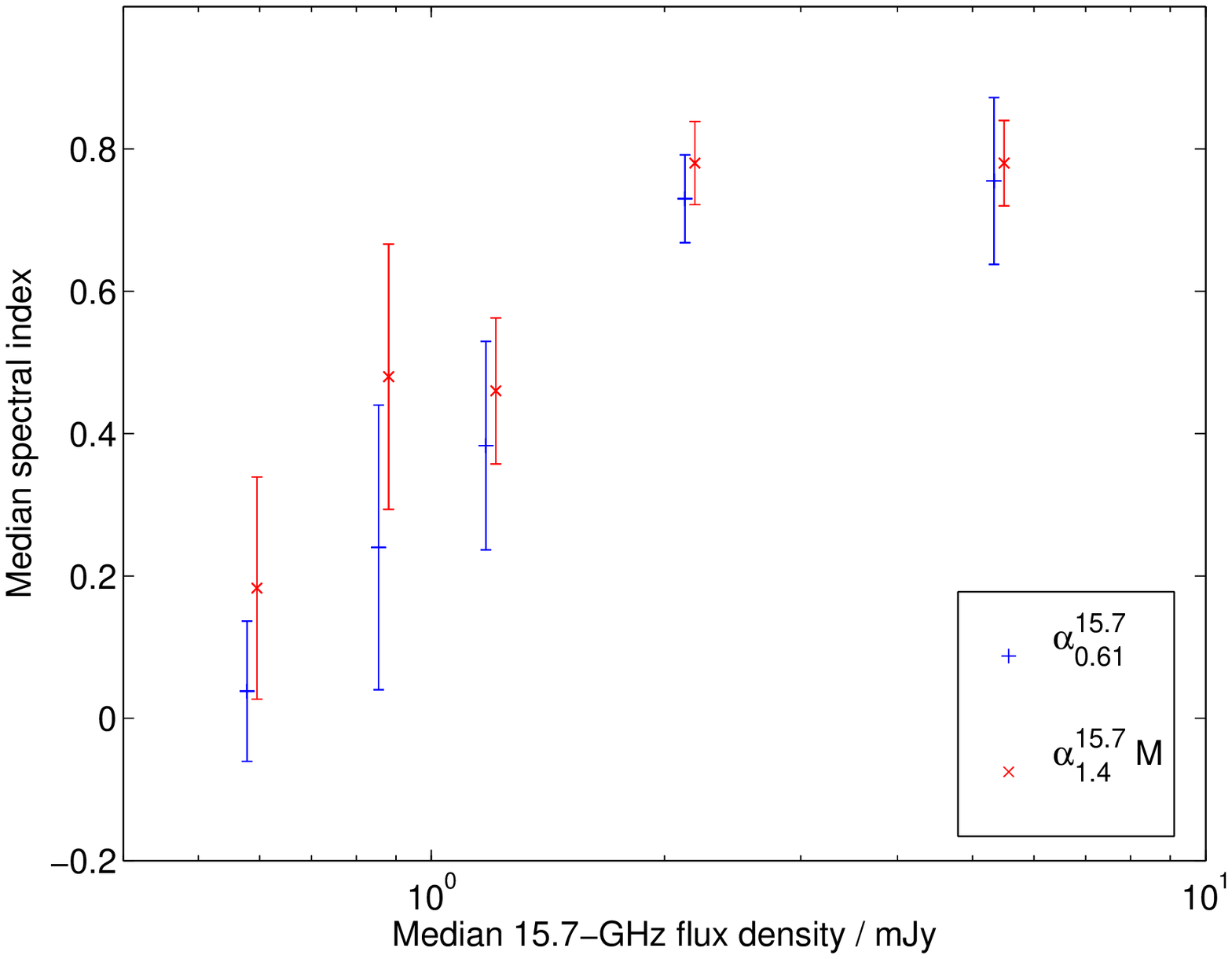}
\qquad
\includegraphics[width=8cm]{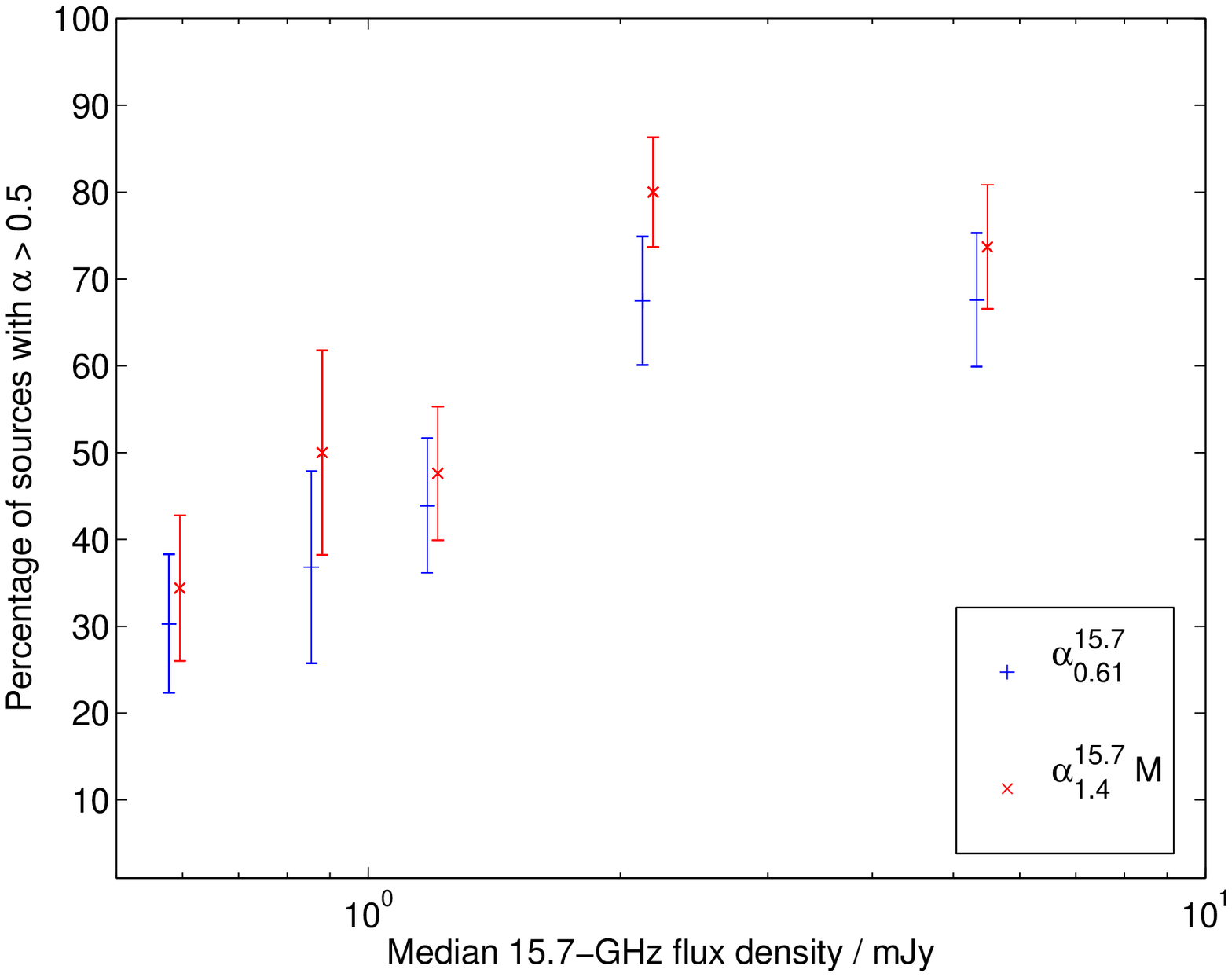}}
\caption{The median spectral index and percentage of sources with $\alpha > 0.5$ as a function of 15.7-GHz flux density for sources in the complete samples A and B. The values of $\alpha^{15.7}_{1.4}M$ are calculated using the best 1.4-GHz flux density values available (order of preference = NVSS, WSRT, OM2008, FIRST, BI2006). The points are plotted at the median flux density in each bin but are offset for clarity. The sources in the lowest two flux density bins are from Sample A, which is complete above 0.5~mJy and the remaining three flux density bins contain sources from sample B, complete above 1~mJy. Limiting spectral indices are included in the median using survival analysis (see text). The error bars in the median spectral index plot show the interquartile range divided by $\sqrt{N}$ to give an indication of the errors in the medians. \label{fig:median-complete}}

\end{figure*}

\begin{table*}
\caption{The $\alpha_{0.61}^{15.7}$ results for three 15.7-GHz flux density bins.}\label{tab:alpha610}
\smallskip
\begin{center}
\begin{tabular}{p{1cm}p{2.7cm}p{1.2cm}p{1.5cm}p{1.2cm}p{1.5cm}p{1.3cm}}\hline
Bin name & 15.7-GHz flux density range / mJy & Number of sources & Number of upper limits & Median $\alpha$ & Mean $\alpha$ & \% $\alpha$ $>$ 0.5 \\ \hline
$S_{\rm{low}}$  & $0.300 < S \leq 0.755$  &  99 & 45 & 0.08 & $0.25 \pm 0.04$ & $25 \pm 4$ \\
$S_{\rm{med}}$  & $0.755 < S \leq 1.492$  &  97 & 29 & 0.36 & $0.31 \pm 0.05$ & $40 \pm 5$ \\
$S_{\rm{high}}$ & $1.492 < S \leq 45.700$ &  95 & 4  & 0.75 & $0.57 \pm 0.05$ & $67 \pm 5$ \\ \hline
\end{tabular}
\end{center}
\end{table*}

\begin{table*}
\caption{The $\alpha_{1.4}^{15.7} M$ results for three 15.7-GHz flux density bins.}\label{tab:alpha14}
\smallskip
\begin{center}
\begin{tabular}{p{1cm}p{2.7cm}p{1.2cm}p{1.5cm}p{1.2cm}p{1.5cm}p{1.3cm}cp{1.2cm}p{1.2cm}p{1.2cm}p{1.5cm}p{1.3cm}}\hline
Bin name & 15.7-GHz flux density range / mJy & Number of sources & Number of upper limits & Median $\alpha$ & Mean $\alpha$ & \% $\alpha$ $>$ 0.5\\ \hline
$S_{\rm{low}}$  & $0.300 < S \leq 0.755 $ &  99 & 27 & 0.10 & $0.11 \pm 0.07$ & $29 \pm 4$\\
$S_{\rm{med}}$  & $0.755 < S \leq 1.492 $ &  99 & 13 & 0.43 & $0.30 \pm 0.07$ & $46 \pm 5$\\
$S_{\rm{high}}$ & $1.492 < S \leq 45.700$ &  98 & 2  & 0.79 & $0.66 \pm 0.05$ & $76 \pm 4$\\ \hline
\end{tabular}
\end{center}
\end{table*}

\begin{figure}
\centerline{
\includegraphics[width=8.2cm]{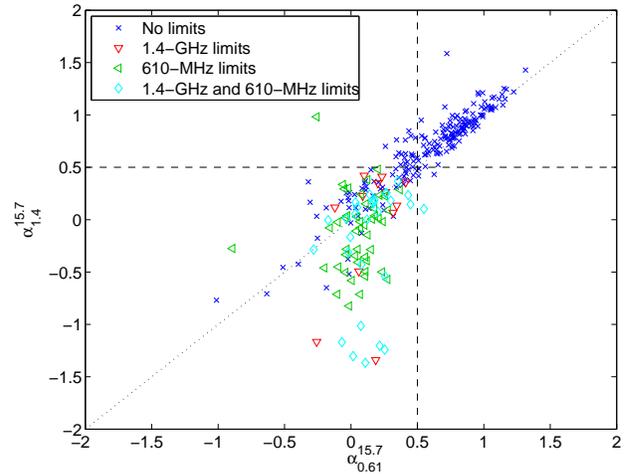}}
\caption{Values of $\alpha^{15.7}_{0.61}$ and $\alpha^{15.7}_{1.4}M$, showing which values are upper limits. The arrows indicate the direction in which the points which are upper limits could move. The dashed lines at $\alpha = 0.5$ illustrate the division between steep and flat spectrum sources; sources with $\alpha > 0.5$ are classified as steep spectrum while sources with $\alpha < 0.5$ are classified as flat spectrum. The dotted line illustrates $\alpha^{15.7}_{0.61} = \alpha^{15.7}_{1.4}M$. \label{fig:alpha-alpha}}

\end{figure}

\begin{figure*}
\centerline{\includegraphics[width=8.2cm]{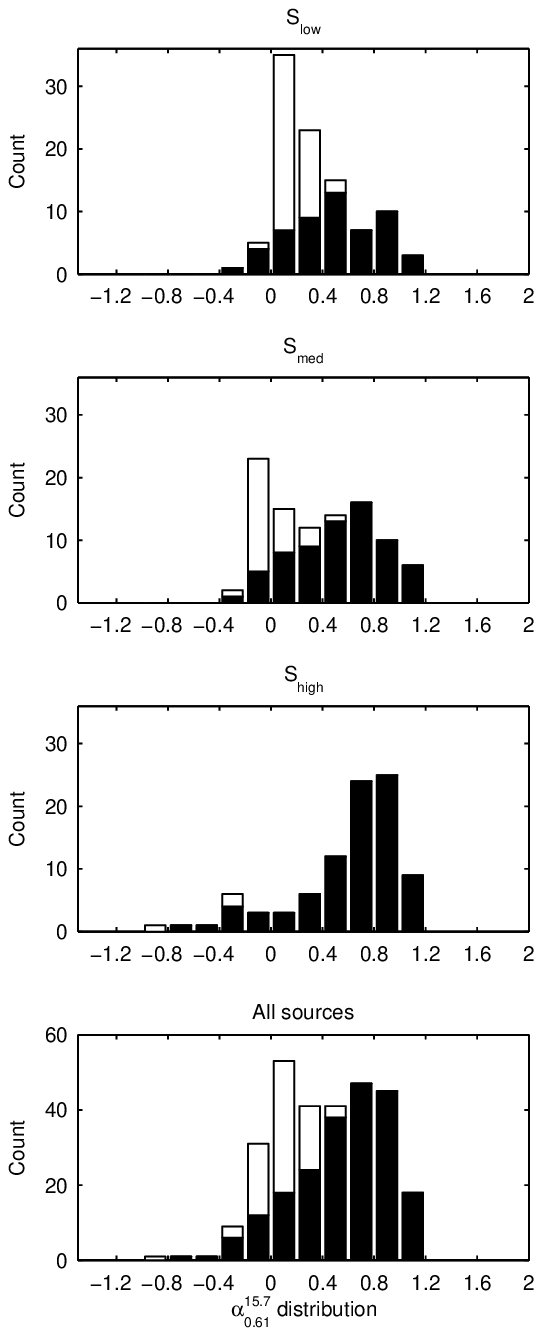} \quad
            \includegraphics[width=8.2cm]{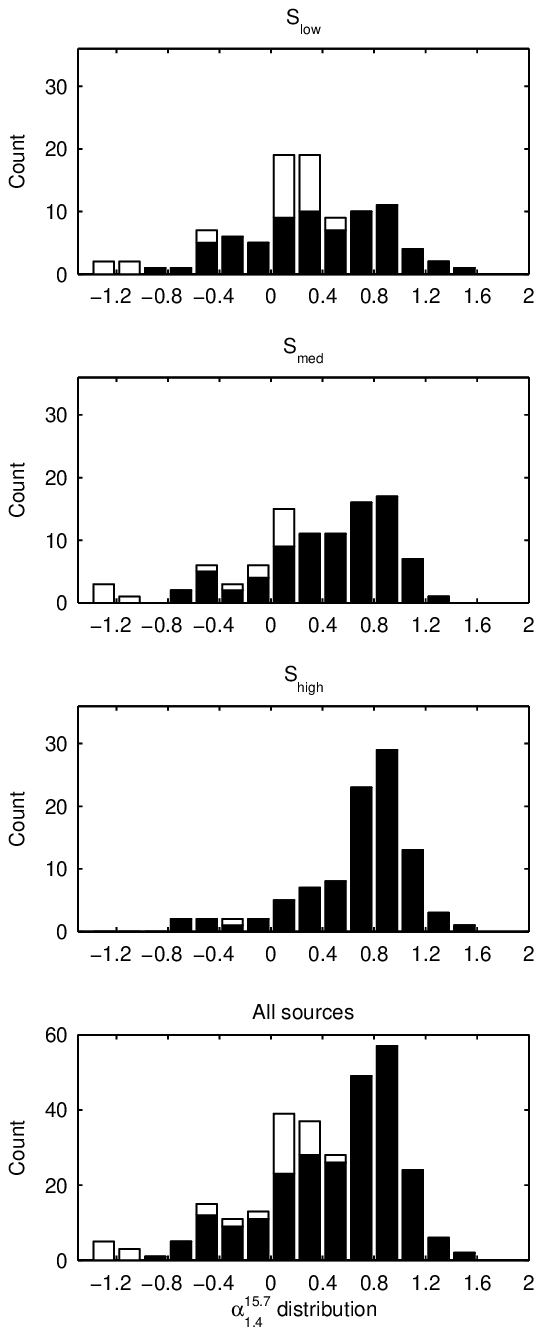}}

\caption{The spectral index distribution for three different flux density bins are shown in the top three panels. $\rm{S_{low}}: 0.300 < \rm{S_{15.7 GHz}} \leqslant 0.755$ mJy, $\rm{S_{med}}: 0.755 < \rm{S_{15.7 GHz}} < 1.492$ mJy, $\rm{S_{high}}: 1.492 < \rm{S_{15.7 GHz}} \leqslant 47$ mJy. The spectral index distribution for the whole 10C sample is shown in the bottom panel. $\alpha_{0.61}^{15.7}$ is calculated using the GMRT values and limits, accounting for the resolution difference as described in Section \ref{section:flux-values}, and using the integrated flux density from the 10C catalogue. $\alpha_{1.4}^{15.7}$ was calculated using the best 1.4~GHz flux density available (order of preference = NVSS, WSRT, OM2008, FIRST, BI2006). In both cases upper limits on $\alpha$ are included for sources where no low frequency flux density data is available; these are shown in white while values are plotted in black. \label{fig:spectradist}}
\end{figure*}

Some example radio spectra are shown in Fig. \ref{fig:spectra}, demonstrating the different spectral types observed. Spectral indices are calculated for all sources, as described in Section \ref{section:alpha-methods}. The spectral index $\alpha^{15.7}_{1.4}M$ against 15.7-GHz flux density for all sources is shown in Fig. \ref{fig:alpha-S}. It is clear from this plot that there is a greater proportion of flat spectrum sources at lower flux densities. This trend is further investigated by calculating the median spectral indices $\alpha^{15.7}_{0.61}$ and $\alpha^{15.7}_{1.4}M$ in three different flux density bins containing equal numbers of sources and the results are shown in Tables \ref{tab:alpha610} and \ref{tab:alpha14}. Upper limits are included by using the ASURV Rev 1.2 package which implements the survival analysis methods presented in \citet{1985ApJ...293..192F}.  The distributions of both $\alpha^{15.7}_{0.61}$ and $\alpha^{15.7}_{1.4}M$ as a function of 15.7-GHz flux density show a sharp change at flux densities between 1 and 2 mJy. The spectral index distributions for these three flux density bins for both $\alpha^{15.7}_{0.61}$ and  $\alpha^{15.7}_{1.4}M$ are shown in Fig. \ref{fig:spectradist} and are very similar. There is a distinct peak at $\alpha \sim 0.7$ in the highest flux density bin ($\rm{S_{high}}$). As flux density decreases the peak broadens as the contribution from flat spectrum sources becomes much more significant. In the lowest flux density bin ($\rm{S_{low}}$) the sources display a wide range of spectral index values, with a broad peak at $\alpha \sim 0.3$. The sources shown in white are upper limits and could only move to the left in these plots, making the $\rm{S_{low}}$ distribution even more different from the $\rm{S_{high}}$ distribution.

 This is illustrated in Fig. \ref{fig:median-all} which shows the median $\alpha^{15.7}_{0.61}$, $\alpha^{15.7}_{1.4}M$ and $\alpha^{15.7}_{1.4}N$ and the percentage of sources with $\alpha > 0.5$ in narrower 15.7-GHz flux density bins -- there is a far higher proportion of flat spectrum sources at lower flux densities. The values for $\alpha^{15.7}_{1.4}N$  (using NVSS flux densities only) are very similar to those for $\alpha^{15.7}_{1.4}M$, except in the lowest flux density bin which is dominated by upper limits in the NVSS-only case. This implies that resolution differences between 10C and the other surveys is not having a major effect on the derived spectral indices.  The same plots for the complete samples A and B are shown in Fig. \ref{fig:median-complete}. There are fewer sources in this sample so the uncertainties are larger but the general trend towards decreasing spectral indices below $\approx$ 2~mJy remains the same.  

A plot of $\alpha^{15.7}_{1.4}M$ against $\alpha^{15.7}_{0.61}$ and for all sources is shown in Fig. \ref{fig:alpha-alpha}. There is a good correlation between the two values of spectral index. The majority of the points in the bottom left corner which deviate from the correlation are upper limits and therefore could move closer to the one-to-one correlation line shown. Of the points which are not upper limits, slightly more lie above the one-to-one correlation line than below it, indicating a slight spectral steepening at higher frequencies.

Possible effects of variability are discussed in Section \ref{section:variability}.  The variability properties of the population studied here are not known, but it seems likely that, if anything, variability would increase the proportion of steep spectrum sources rather than producing the decrease observed. It is possible, for example, that variability in some of the genuinely flat spectrum sources over the epochs covered by the surveys used here could make them appear to have steep spectra. On the other hand, the genuinely steep spectrum sources in our faint sample are unlikely to be variable (unless their properties differ markedly from the steep spectrum sources at higher flux densities) so their spectral indices will not be affected at all.

\subsection{Comparison with other spectral index studies}\label{section:alphadisc}

\begin{figure}
\centerline{\includegraphics[width=8.2cm]{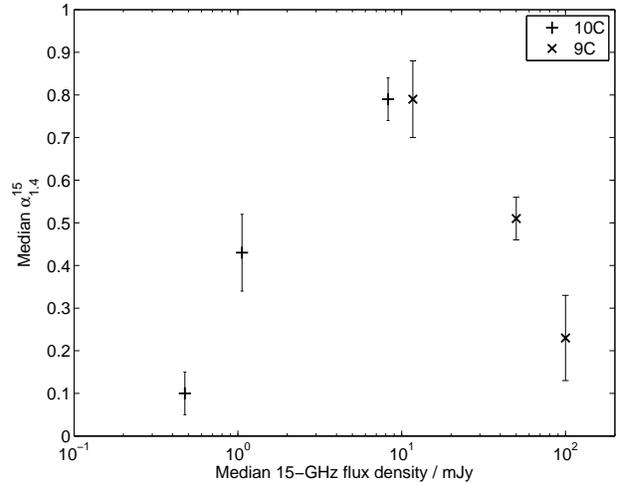}}
\caption{The median spectral index as a function of flux density from the study of 9C sources by \citet{2010MNRAS.404.1005W} and the 10C sources studied in this paper. The points are plotted at the bin midpoint, except for the highest flux density bin ($S > 100$ mJy) where the point is plotted at the bin lower limit.}\label{fig:10C-9C}
\end{figure}

The variation in spectral index with flux density at higher ($\sim$10 mJy -- 1 Jy) flux densities has been investigated by \citet{2010MNRAS.404.1005W} and \citet{2011MNRAS.412..318M}. \citeauthor{2010MNRAS.404.1005W} and \citeauthor{2011MNRAS.412..318M} used samples selected at 15 and 20 GHz respectively and, by matching them to catalogues at 1.4~GHz, found that the median spectral index becomes rapidly larger with decreasing flux density; for example Waldram et al. found that the median $\alpha^{15}_{1.4}$ changed from around 0 at 1~Jy to 0.8 at 10~mJy. Our study of 10C sources shows that as the flux density decreases further the median spectral index drops again, with the median $\alpha^{15}_{1.4}$ decreasing from around 0.8 for flux densities $>1.5$ mJy to around 0.1 at 0.5~mJy. As discussed in Section \ref{section:spectral} the change occurs fairly abruptly at $\approx 1$mJy, indicating that the nature of the 15-GHz source population is changing at this flux density. The relationship between spectral index and flux density is summarised in Fig. \ref{fig:10C-9C} which shows the \citeauthor{2010MNRAS.404.1005W} results from the Ninth Cambridge survey (9C) along with the results from this study. 

We have shown here that the spectral index distribution changes dramatically below $\sim$1~mJy for a sample of sources selected at 15.7~GHz. A similar trend has been found by other studies at slightly lower frequencies. For example, \citet{2006A&A...457..517P} studied a complete sample of 131 sources detected at 5 and 1.4~GHz with a comparable flux density range to the sample studied here. They found that the median spectral index between 5 and 1.4 GHz changed from $0.56 \pm 0.06$ for sources with $S_{5 \rm{ GHz}} > 4$~mJy to $0.24 \pm 0.06$ for sources with $S_{5 \rm{ GHz}} < 4$~ mJy.

\subsection{Extent of the radio emission}\label{section:extended-results}

\begin{figure}
\centerline{\includegraphics[width=8.2cm]{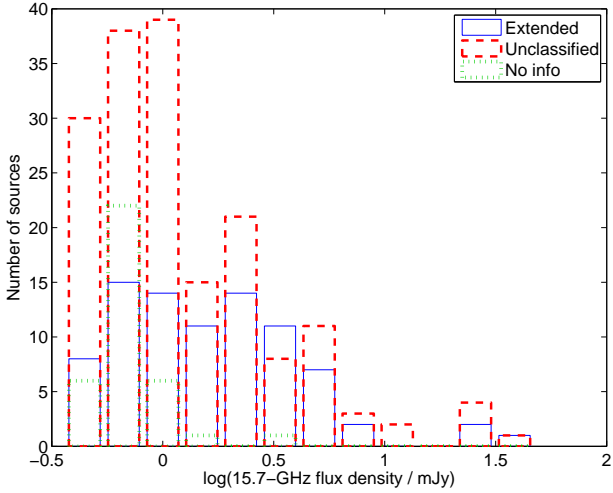}}
\caption{The 15.7-GHz flux density distribution for the extended sources, those which are unclassified and those with no information. The three histograms are overlaid.}\label{fig:flux-extended}
\end{figure}

The sources were split into three groups (extended, unclassified and no information) as described in Section \ref{section:extended-methods}. A summary of the properties of these three groups of sources in terms of spectral index and flux density is given in Table \ref{tab:extended} and the flux density distribution for the three groups is shown in Fig. \ref{fig:flux-extended}. A larger proportion of the brighter sources are extended, with extended sources making up 38 percent of the total number of sources with $S > 1$~mJy, compared to 19 percent of sources with $S < 1$~mJy. This could be a result of the more stringent criteria for classifying sources with low signal to noise as extended as well as the fact that extended low surface brightness emission is more likely to be missed at low signal to noise levels. As expected, the majority of the sources with no information are faint, with $S < 1$~mJy. 

\begin{table}
\caption{A summary of the properties of sources classified as extended, those not classified and those with no information. The value of spectral index used here is $\alpha^{15.7}_{0.61}$ so the five sources which are outside the GMRT survey area are not included. The numbers in brackets refer to the percentage of sources in each classification.}\label{tab:extended}
\medskip
\begin{center}
\begin{tabular}{cccccc}\hline
    &  \multicolumn{5}{c} {Number of sources with ...}\\
Bin          & Total & $S_{15.7} > 1$~mJy & $S_{15.7} < 1$~mJy & $\alpha > 0.5$ & $\alpha < 0.5$ \\\hline
Extended     & 85    & 55 (65) & 30 (35) & 70 (82) & 15  (18)\\
Unclassified & 170   & 80 (47) & 90 (53) & 56 (33) & 114 (67)\\
No info      & 36    &  5 (14) & 31 (86) & 2  (6)  & 34  (94)\\\hline
\end{tabular}
\end{center}
\end{table}

\begin{figure}
\centerline{\includegraphics[width=8.2cm]{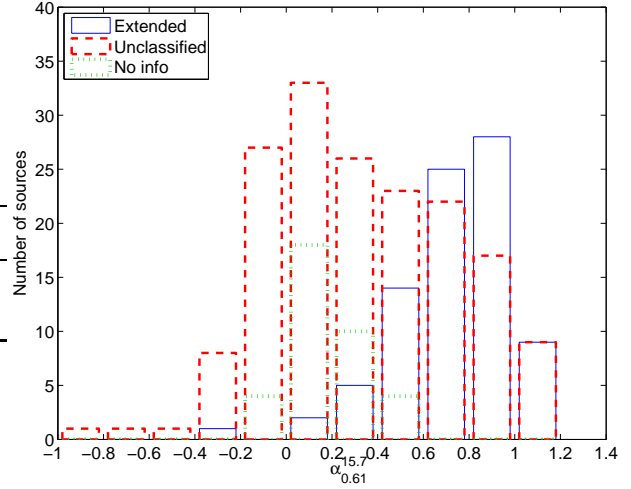}}
\caption{The spectral index distribution for sources classified as extended and unclassified, and those with no information. The three histograms are overlaid. Upper limits in $\alpha$ are plotted for the sources with no information which explains the apparent peak at $\alpha \approx 0$.}\label{fig:alpha-extended}
\end{figure}

\begin{figure}
\centerline{\includegraphics[width=8.2cm]{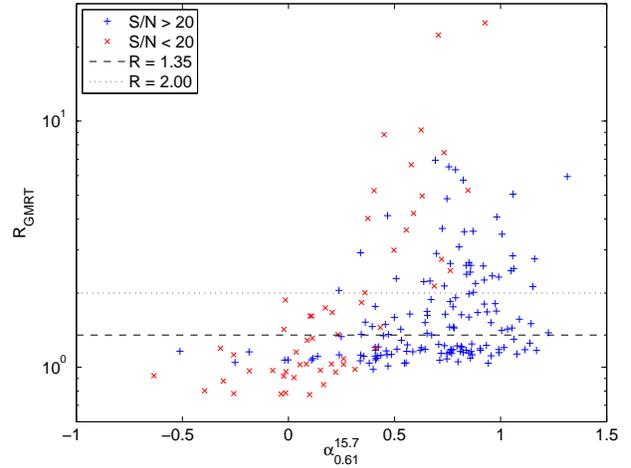}}
\caption{$R_{GMRT}$ as a function of $\alpha^{15.7}_{0.61}$ for all sources in the GMRT area. Sources with peak flux / noise $<$ 20 in the GMRT catalogue are shown separately and the horizontal lines at $R = 1.35$ and 2.00 indicate the cutoffs in $R$ used when classifying the extended sources, see text for details. }\label{fig:alpha-RGMRT}
\end{figure}

The spectral index distributions for the extended sources and those not classified are, as anticipated, significantly different, as shown in Fig. \ref{fig:alpha-extended} and Table \ref{tab:extended}. The majority (70/85) of the extended sources are steep spectrum, with the spectral index distribution peaking at $\alpha \approx 0.8$. The distribution of the unclassified sources is much broader and is less peaked, with the distribution stretching between $\alpha \approx -0.2$ to $\alpha \approx +1$. The majority of the sources with no information display a flat spectrum, as these are the sources which are too faint to be detected at the lower frequencies of 610~MHz/1.4~GHz. For these sources the value of $\alpha$ plotted is an upper limit, which explains the large peak at $\alpha \approx 0$. Fig. \ref{fig:alpha-RGMRT} shows $R_{GMRT}$ against $\alpha^{15.7}_{0.61}$ for all sources matched to the GMRT catalogue. It is evident from this plot that, as expected, a greater number of the extended sources have a steep spectrum.

\begin{figure*}
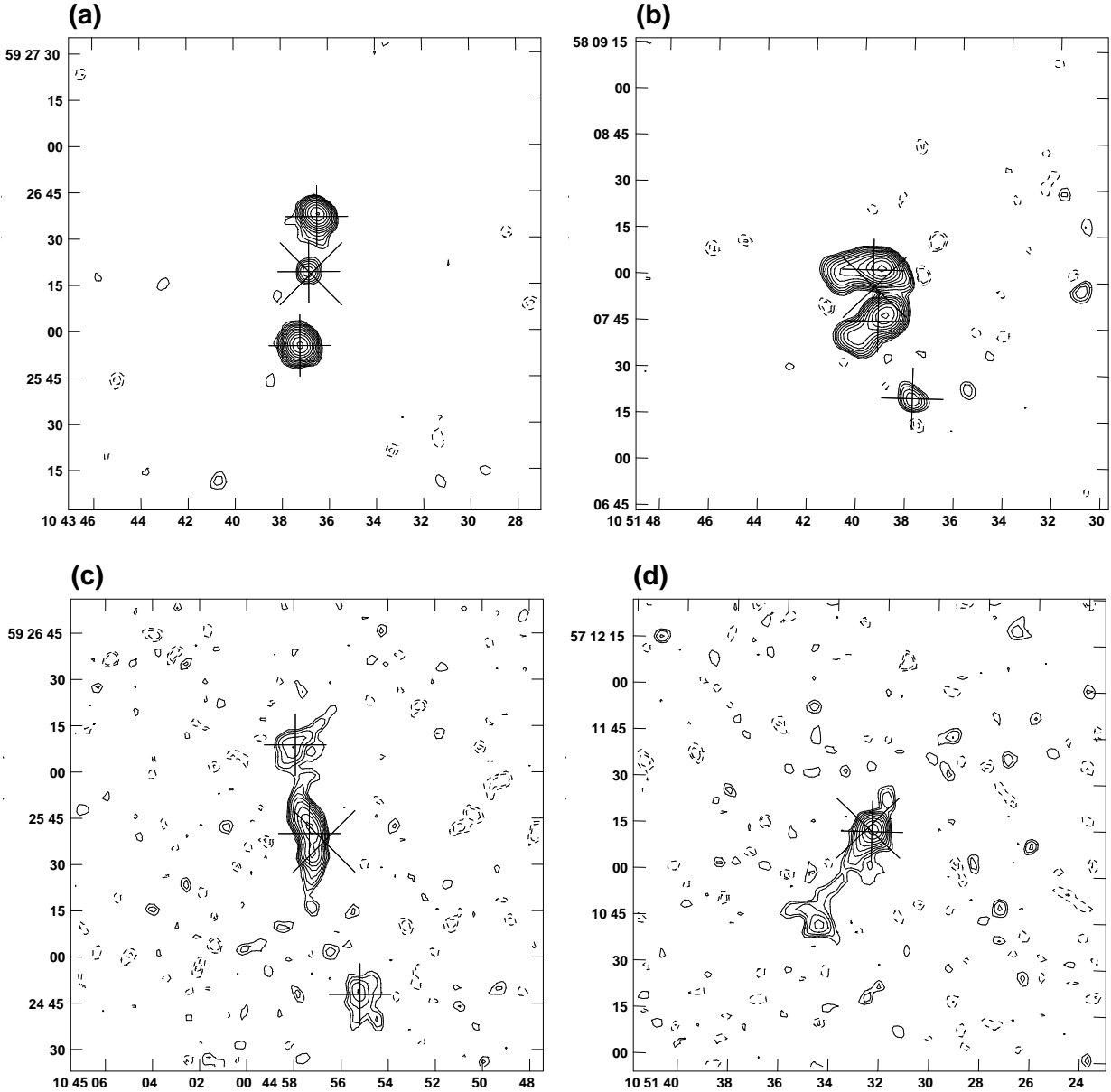

\centerline{\includegraphics[bb=54 214 558 706,clip=,width=8cm]{10CJ104336+592618.eps} \quad
            \includegraphics[bb=54 214 558 706,clip=,width=8cm]{10CJ105139+580757.eps}}
\bigskip

\centerline{\includegraphics[bb=54 214 558 706,clip=,width=8cm]{10CJ104456+592537.eps}\quad
            \includegraphics[bb=54 214 558 706,clip=,width=8cm]{10CJ105132+571114.eps}}
\caption{GMRT images of some example extended sources. The contours are drawn at $(\pm 2\sqrt{2^n}, n = 0, 1... 7) \times x \textrm{ mJy}$ where $x$ = 0.16 for (a), 0.15 for (b), 0.086 for (c) and 0.076 for (d). The $\times$ marks the position of the 10C source and the + mark the positions of the components listed in the GMRT catalogue for each source. Sources (a) and (b) are steep spectrum sources with $\alpha^{15.7}_{0.61} = 0.93$ and 1.08 respectively. Sources (c) and (d) have flatter spectra with $\alpha^{15.7}_{0.61} = 0.49$ and 0.34 respectively.\label{fig:extended}}
\end{figure*}

The distribution for extended sources displays a flat spectrum tail, 15 sources having $\alpha^{15.7}_{0.61} < 0.5$. It is likely that these extended, flat spectrum sources are extended sources with a flat spectrum core which dominates at high frequency. We might therefore expect that the lower frequency images of most of these sources would display a dominant core surrounded by some, possibly fainter, extended emission, whereas the steep spectrum extended sources would have relatively more pronounced lobes. Examination of the images shows this to be the case for most sources; typical examples of steep and flat spectrum extended sources are shown in Fig. \ref{fig:extended}.

\subsection{Correlations between size and spectral index}

In Section \ref{section:extended-results} it is shown that the majority of the flat spectrum sources are not classified as extended and therefore probably compact. However, a small but significant number (15 out of 163) of the flat spectrum sources are clearly extended, indicating a variety of source types. There are also a significant number of sources (34) for which we have no information about their structure. 

Recent work by \citet{2011MNRAS.412..318M} using the AT20G survey found that for $\alpha^5_1$ there was a clear divide between the extended, steep spectrum sources and the compact, flat spectrum sources. However, when looking at the higher frequency spectral index, $\alpha^{20}_8$, they found that the extended sources displayed a much broader range in spectral index. This is consistent with the flat spectrum tail in the spectral index distribution for extended sources observed here and supports the idea that a flat spectrum core is becoming increasingly dominant at higher frequencies. A study by \citet{2006A&A...457..517P} investigated the properties of a sample of 111 sources using data at 1.4 and 5~GHz and found that nearly all the extended or multiple sources were steep spectrum. The lack of flat spectrum extended sources in the \citeauthor{2006A&A...457..517P} study is probably due to the lower frequencies used, as at these frequencies the flat spectrum core has not yet become dominant.
 
\subsection{Effect on the source counts}\label{section:counts}

The 15.7-GHz source count derived from the full 10C survey is presented in \citet{2011MNRAS.415.2708D}. The function fitted to the source count is a broken power law, as shown in equation \ref{eqn:sourcecount},

\begin{equation}
n(S) \equiv \frac{dN}{dS} \approx \left\{
 \begin{array}{l l}
    24 \left(\frac{S}{\rm{Jy}}\right)^{-2.27} \textrm{ Jy}^{-1} \textrm{ sr}^{-1} &  \textrm{ for 2.8 $\leq S \leq$ 25 mJy},\\
    376 \left(\frac{S}{\rm{Jy}}\right)^{-1.80} \textrm{ Jy}^{-1} \textrm{ sr}^{-1} & \textrm{ for 0.5 $\leq S <$ 2.8 mJy}.\\
  \end{array} \right.
\label{eqn:sourcecount}
\end{equation}  

It is significant that the break in this power law occurs at 2.8~mJy i.e. at approximately the same flux density as the change in the spectral index distribution observed here.

To examine this further, the source counts for the complete sample of sources (made up of samples A and B) studied in this paper are presented in Table \ref{tab:counts}. The counts for steep and flat spectrum sources are shown separately; the flux density bins are double the bins used in presenting the full 10C counts in \citet{2011MNRAS.415.2708D}. No attempt is made to fit the source count due to the small number of sources in each bin. However, it is clear that the source counts are significantly different for the two populations, with, as expected, a greater proportion of flat spectrum sources in the fainter flux density bins ($S_{15\rm{ GHz}} \lesssim 1$~mJy), while the steep spectrum sources dominate at the higher flux densities.  Taken with the flattening of the overall counts for flux densities $< 2.8$~mJy, this indicates that the source counts for the steep spectrum sources must be flattening significantly at flux densities below about 1~mJy.

\begin{table}
\caption{Source counts for the complete 10C sample from the Lockman Hole fields (made up of samples A and B). Steep and flat spectrum sources (with $\alpha^{15.7}_{0.61} > 0.5$ and $\alpha^{15.7}_{0.61} < 0.5$ respectively) are shown separately. Each flux density bin corresponds to two of the flux density bins used in presenting the full 10C source counts in \citet{2011MNRAS.415.2708D} (except for the highest flux density bin).}\label{tab:counts}

\begin{center}
\begin{tabular}{ddddd}\hline
\dhead{Bin start} & \dhead{Bin end} & \dhead{No. of sources}   & \dhead{No. of sources}   & \dhead{Area}\\
\dhead{/mJy}      & \dhead{/mJy}    & \dhead{with $\alpha > 0.5$} & \dhead{with $\alpha < 0.5$} & \dhead{/deg$^2$}\\\hline
9.000 & 25.000 & 4  & 4  & 4.64\\
2.900 & 9.000  & 22 & 11 & 4.64\\
1.500 &	2.900  & 27 & 10 & 4.64\\
1.000 &	1.500  & 18 & 24 & 4.64\\
0.775 &	1.000  & 5  & 11 & 1.73\\
0.600 &	0.775  & 5  & 11 & 1.73\\
0.500 &	0.600  & 6  & 12 & 1.73\\\hline

\end{tabular}
\end{center}
\end{table}

There is evidence from other studies that the FRII population is dropping out at around the flux density where we observe the flattening of the steep spectrum source count.  For example, \citet{2008MNRAS.390..819G} used a sample of Combined NVSS-FIRST Galaxies (CoNFIG) to construct a 1.4-GHz source count for FRI and FRII sources. They found that the FRII population is dropping out as flux density decreases below approximately 20~mJy at 1.4~GHz, leaving a source population dominated by FRI sources at lower flux densities. This change in the population at $S_{1.4\rm{ GHz}} \approx 20$~mJy corresponds to $S \approx 3$~mJy at 15~GHz for a steep spectrum source ($\alpha = 0.75$). The simulated source counts produced by \citet{2008MNRAS.388.1335W} (which are discussed in more detail in Section \ref{section:s3}) also show that the FRII population drops out at a few mJy at 18~GHz. It therefore seems likely that the changes in the spectral index distributions at around 1~mJy in our 15~GHz sample are due in part to the disappearance of the FRII sources.  However, as discussed in Section \ref{section:s3}, the dominance of a flat spectrum population at flux densities below 1~mJy is in clear disagreement with the models of \citeauthor{2008MNRAS.388.1335W}, \citeauthor{2005A&A...431..893D} and \citeauthor{2011A&A...533A..57T}.

\section{Comparison with samples selected at 1.4~GH\lowercase{z}}\label{section:1.4-selected}

The source population at 1.4~GHz has been much more widely studied than the higher frequency population and models of the faint population at higher frequencies are often extrapolated from this lower frequency data. It is therefore useful to see how the spectral index distribution of the 10C source population compares to that for sources selected at 1.4~GHz.

Two samples of sources at 1.4~GHz were used; the first sample was selected from the FIRST catalogue (sample P) and the second from the NVSS catalogue (sample Q). The FIRST catalogue is deeper so sample P provides more information about the faint source population; the NVSS survey on the other hand has a beam size comparable to the 10C survey so provides more reliable spectral indices.

Some consideration needs to be given to the limiting flux densities of the FIRST and NVSS catalogues and the 10C catalogue which they are being matched. A source with a 1.4-GHz flux density of 3.4~mJy (the completeness limit of NVSS) should be detected in 10C unless it has a spectral index greater than 0.8. It is therefore expected that counterparts will be found at 15.7~GHz for the majority of the NVSS sources. FIRST, however, has a completeness limit of 1~mJy; a source with a 1.4-GHz flux density of 1~mJy could be detected at 15.7~GHz if it had $\alpha^{15.7}_{1.4} < 0.3$. We therefore expect a larger proportion of lower limits on the spectral indices for FIRST sources.

\subsection{Sample selected from FIRST}\label{section:FIRST-selected}

All FIRST sources in the 10C deep fields (the areas containing Sample A) were selected (see Fig. \ref{fig:source_map}) giving a sample of 127 sources selected at 1.4~GHz. These sources were matched to the full 10C catalogue using a match radius of 15 arcsec, as described in Section \ref{section:matching}. The difference in resolution between the two catalogues meant that there were several sources which were resolved into multiple components in FIRST but unresolved in 10C. For those cases where several FIRST sources matched to one 10C source, the flux densities of the individual FIRST sources were combined, giving a sample of 105 FIRST sources, 70 of which have a match to 10C. 

\begin{figure}
\centerline{\includegraphics[width=8.2cm]{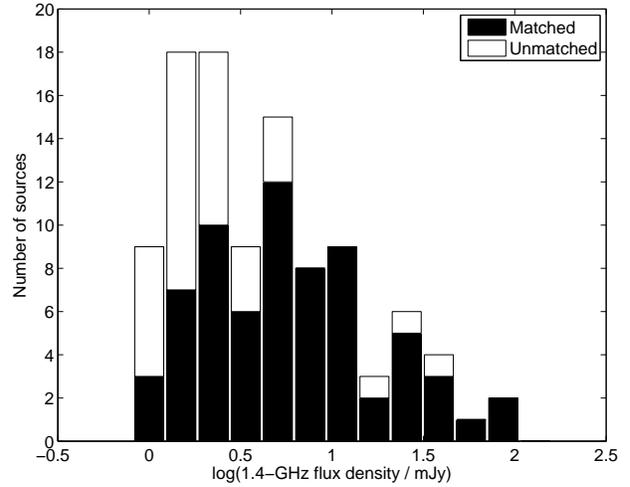}} 
\caption{The flux density distribution of the sources in sample P, selected from the FIRST catalogue. Sources with and without a match to the 10C catalogue are shown separately.}\label{fig:FIRST-flux}
\end{figure}

10C maps of the unmatched FIRST sources were examined by eye. In ten cases, a source was visible in the 10C image -- however its flux was clearly below the 10C completeness limit; the flux density of such a source was found using TVSTAT in AIPS. For the remaining unmatched sources, an upper limit of three times the local noise was placed on the flux density. This allows a lower limit to be placed on the spectral index.

The flux density distribution of the 105 sources in sample P is shown in Fig. \ref{fig:FIRST-flux}.

\subsection{Sample selected from NVSS}\label{section:NVSS-selected}

\begin{figure}
\centerline{\includegraphics[width=8.2cm]{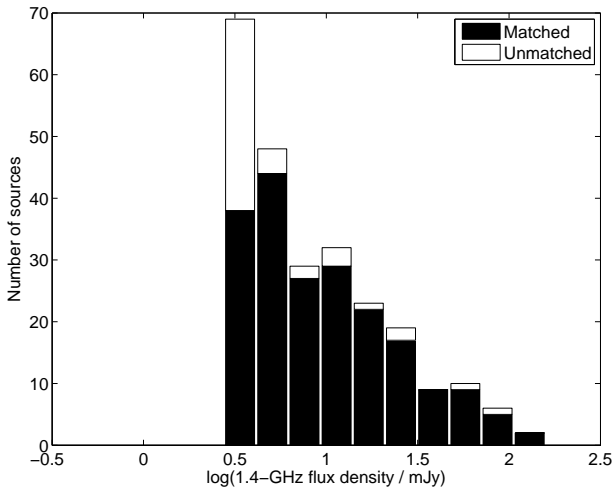}} 
\caption{The flux density distribution of the sources in sample Q, selected from the NVSS catalogue. Sources with and without a match to the 10C catalogue are shown separately.}\label{fig:NVSS-flux}
\end{figure}

In order to have enough NVSS sources to be able to draw statistically significant conclusions, sources were selected from the deep areas of four 10C fields. Two of these are the fields in the Lockman Hole studied in this paper, the other two fields are centred on 17$^{\rm h}$33$^{\rm m}$, +41$^{\circ}$48$^{\rm m}$ and 00$^{\rm h}$24$^{\rm m}$, +31$^{\circ}$52$^{\rm m}$. There are 292 sources in these four fields, giving a sample of comparable size to the sample of 10C sources studied in this paper. The NVSS sources were matched to the 10C catalogue, with limits placed on the unmatched sources as described in Section \ref{section:FIRST-selected}. This sample (sample Q) contains 292 NVSS sources, 223 of which have a match to the 10C catalogue. The flux density distribution of sample Q is shown in Fig. \ref{fig:NVSS-flux}.

\subsection{Spectral index distribution of samples selected at 1.4~GHz}

\begin{figure}
\centerline{\includegraphics[width=8.2cm]{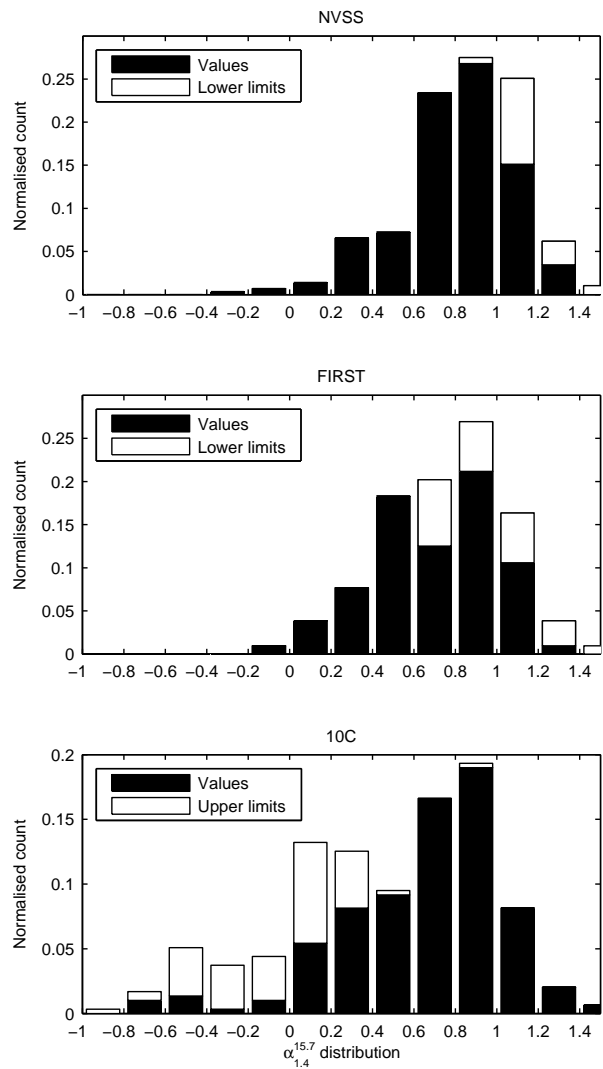}}
\caption{The spectral index distributions for sources selected at 1.4~GHz from NVSS (sample Q, shown in the top panel) and FIRST (sample P, middle panel) and 10C sources selected at 15.7~GHz (bottom panel). Limiting spectral indices for the unmatched sources are shown in white -- note that these are lower limits in the FIRST and NVSS cases, therefore could only move to the right, and upper limits in the 10C case.}\label{fig:alpha-NVSS-FIRST-10C}
\end{figure}

The spectral index $\alpha^{15.7}_{1.4}$ was calculated for all sources in samples P and Q with a match in 10C. For the unmatched sources, a lower limit was placed on the spectral index using the upper limit from the 10C map. Fig. \ref{fig:alpha-NVSS-FIRST-10C} shows a comparison of the spectral index distributions of the sources selected at 15.7~GHz and those selected at 1.4~GHz. As expected the distributions are noticeably different; the sources selected at 1.4~GHz show one peak at $\alpha^{15.7}_{1.4} \approx 0.7$, while the sample selected at 15.7~GHz displays an additional peak at $\alpha^{15.7}_{1.4} \approx 0.3$. The additional population of flat spectrum sources, as expected, are poorly represented by selecting at 1.4~GHz. This is the challenge that extrapolating from lower frequencies to predict the high frequency radio population presents. It relies on accurate modelling of the source population and of how the spectral behaviour of sources varies with frequency; it is particularly unreliable when there are no observations at nearby frequencies at the required flux density. The 10C sample is compared to one such model in Section \ref{section:s3}.

\section{Comparison with the SKADS Simulated Sky}\label{section:s3}

\citet{2008MNRAS.388.1335W,2010MNRAS.405..447W} produced a semi-empirical simulation of the extragalactic radio continuum sky which contains $\approx320$ million sources. This simulation covers a sky area of $20 \times 20$ $\rm deg^2$ out to a cosmological redshift of $z = 20$  and down to flux density limits of 10 nJy at 151, 610~MHz 1.4, 4.86 and 18~GHz. The sources in the simulation are split into six distinct source types: radio-quiet AGN (RQQ), radio loud AGN of the Fanaroff-Riley type I (FRI) and type II (FRII), GHz-peaked spectrum (GPS) sources, quiescent starforming and starbursting galaxies. These simulated sources are drawn from the observed or extrapolated luminosity functions. In order to produce a sub-sample of simulated sources comparable to the sources observed in this work, sources with a flux density greater than 0.5~mJy at 18~GHz were selected from the simulation. This produced a simulated sub-sample of 16235 sources (sample S$^3$). The spectral index between 610~MHz and 18~GHz was calculated from the flux densities in the catalogue.

\begin{figure*}
\centerline{\includegraphics[width=5.5cm]{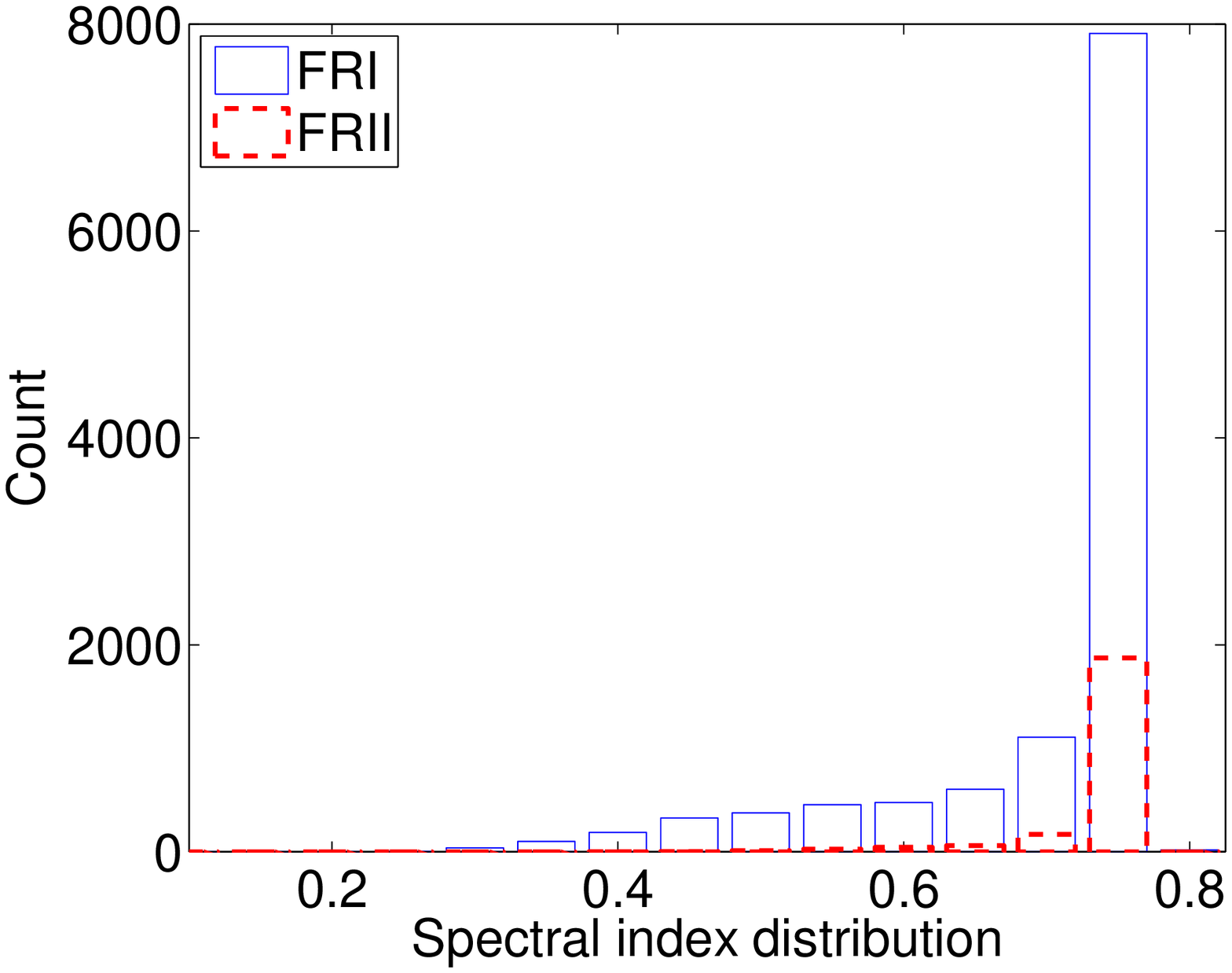}
            \includegraphics[width=5.5cm]{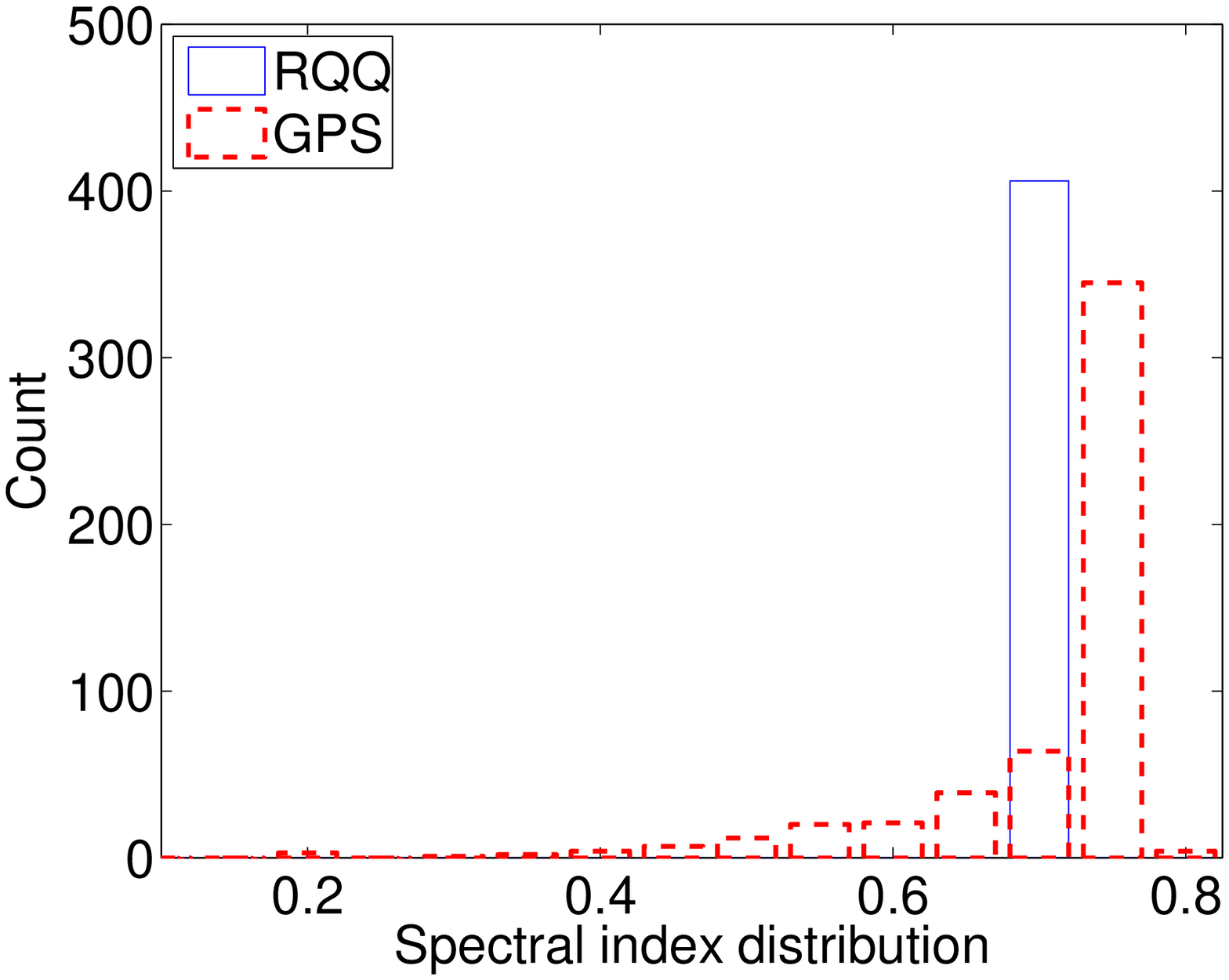}
            \includegraphics[width=5.5cm]{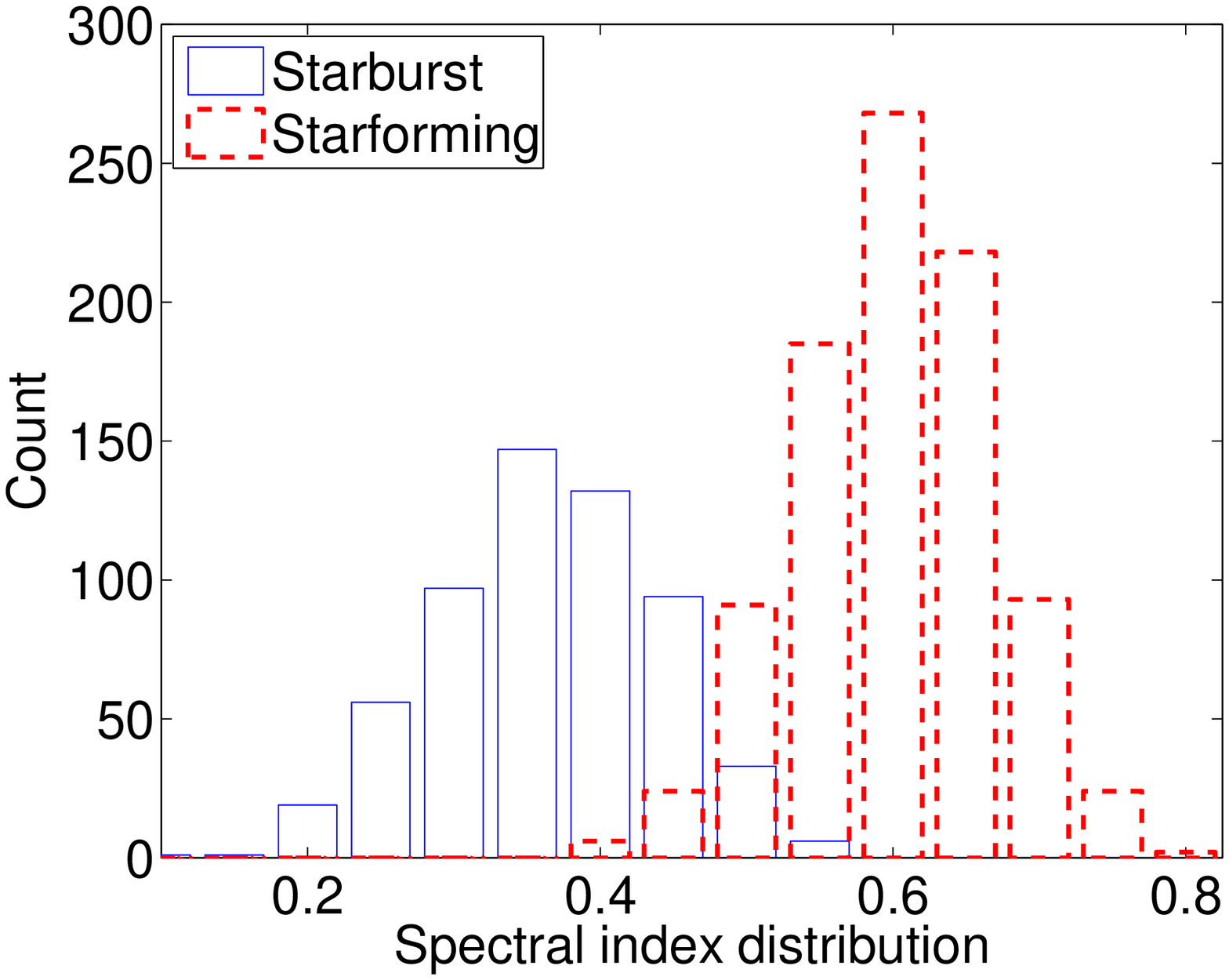}}
\caption{The spectral index distribution of the different source population in the $\rm S^3$ sample with $\rm S_{18~GHz} > 0.5~mJy$. The left panel shows FRI and FRII sources, the middle panel shows radio quiet AGNS and GHz-peaked sources and the right panel shows starburst and quiescent starforming soures.}\label{fig:s3-type-alpha}

\end{figure*}

\begin{figure*}
\centerline{\includegraphics[width=8cm]{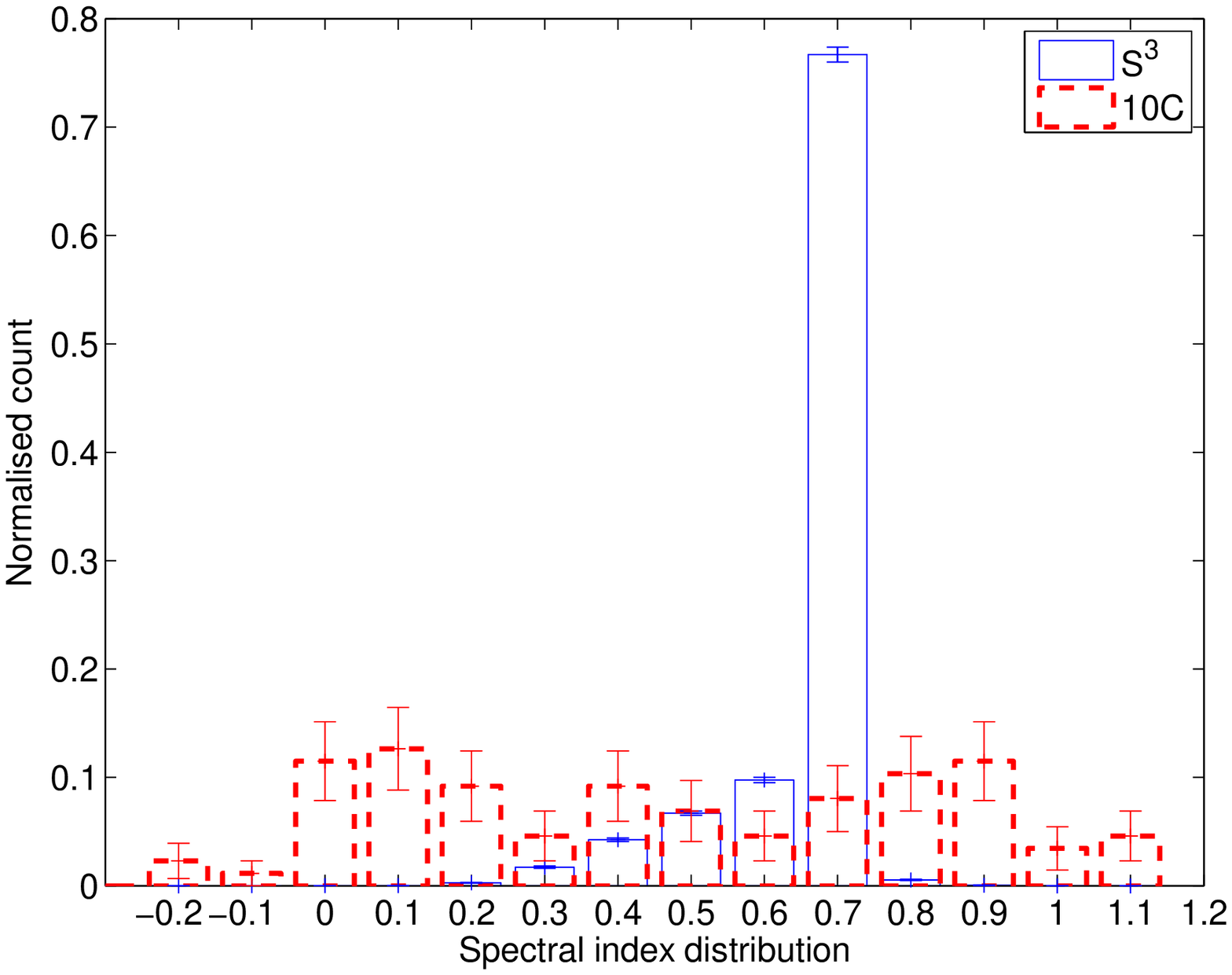}
            \includegraphics[width=8cm]{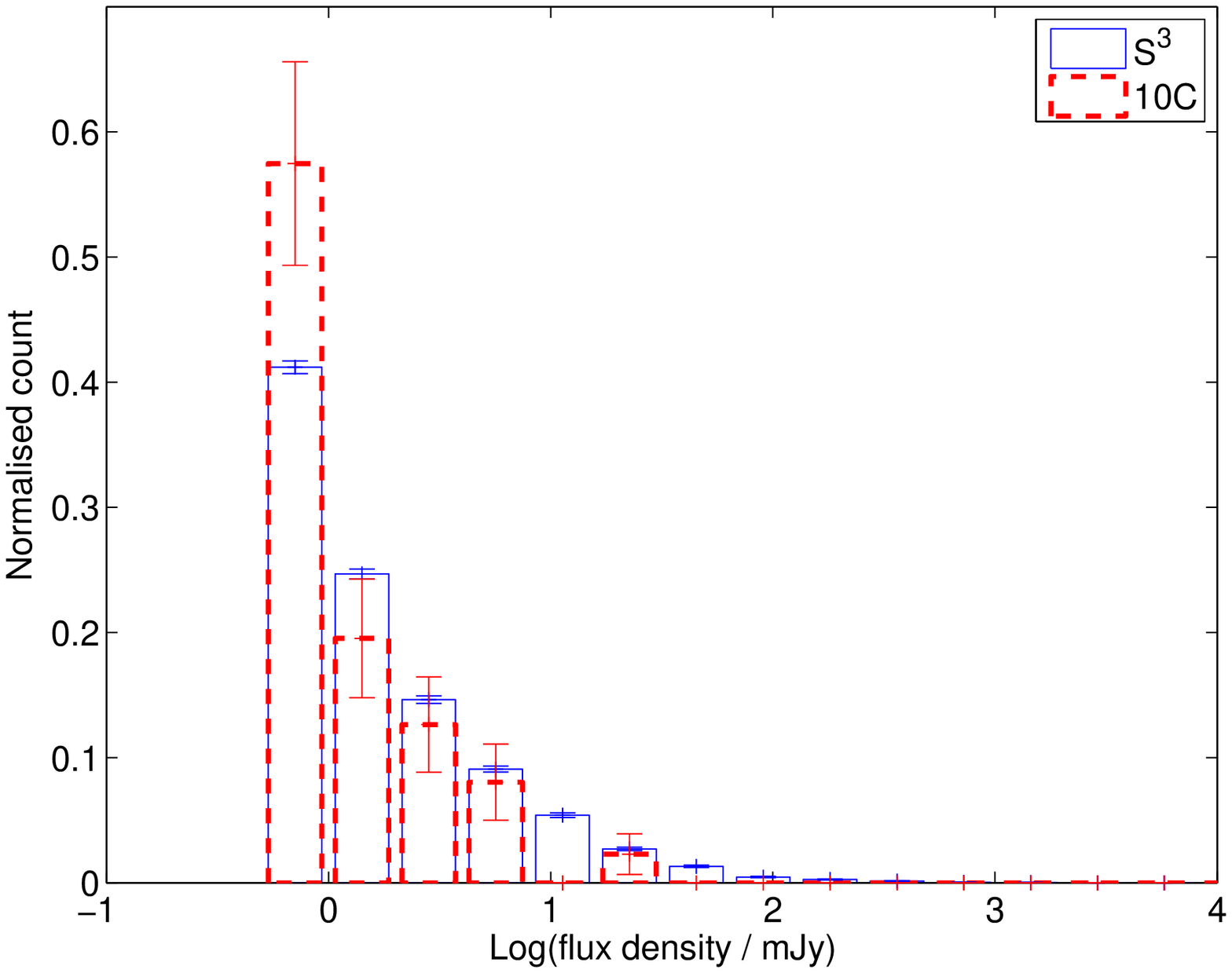}}
\caption{Comparison of the 10C and the simulated samples. The spectral index distribution (left) and flux density distribution (right) for sources in the $\rm S^3$ sample with $\rm S_{18~GHz} > 0.5~mJy$ and for 10C sources with $\rm S_{15.7~GHz} > 0.5~mJy$ is shown. Note that for the $\rm S^3$ sample the values plotted are $\alpha^{18}_{0.61}$ and 18-GHz flux density while for the 10C sources $\alpha^{15.7}_{0.61}$ and 15.7-GHz flux density are plotted. We do not expect this to make any difference to the results. Error bars show the poisson errors.}\label{fig:s-alpha-flux}
\end{figure*}

\begin{figure*}
\centerline{\includegraphics[width=8cm]{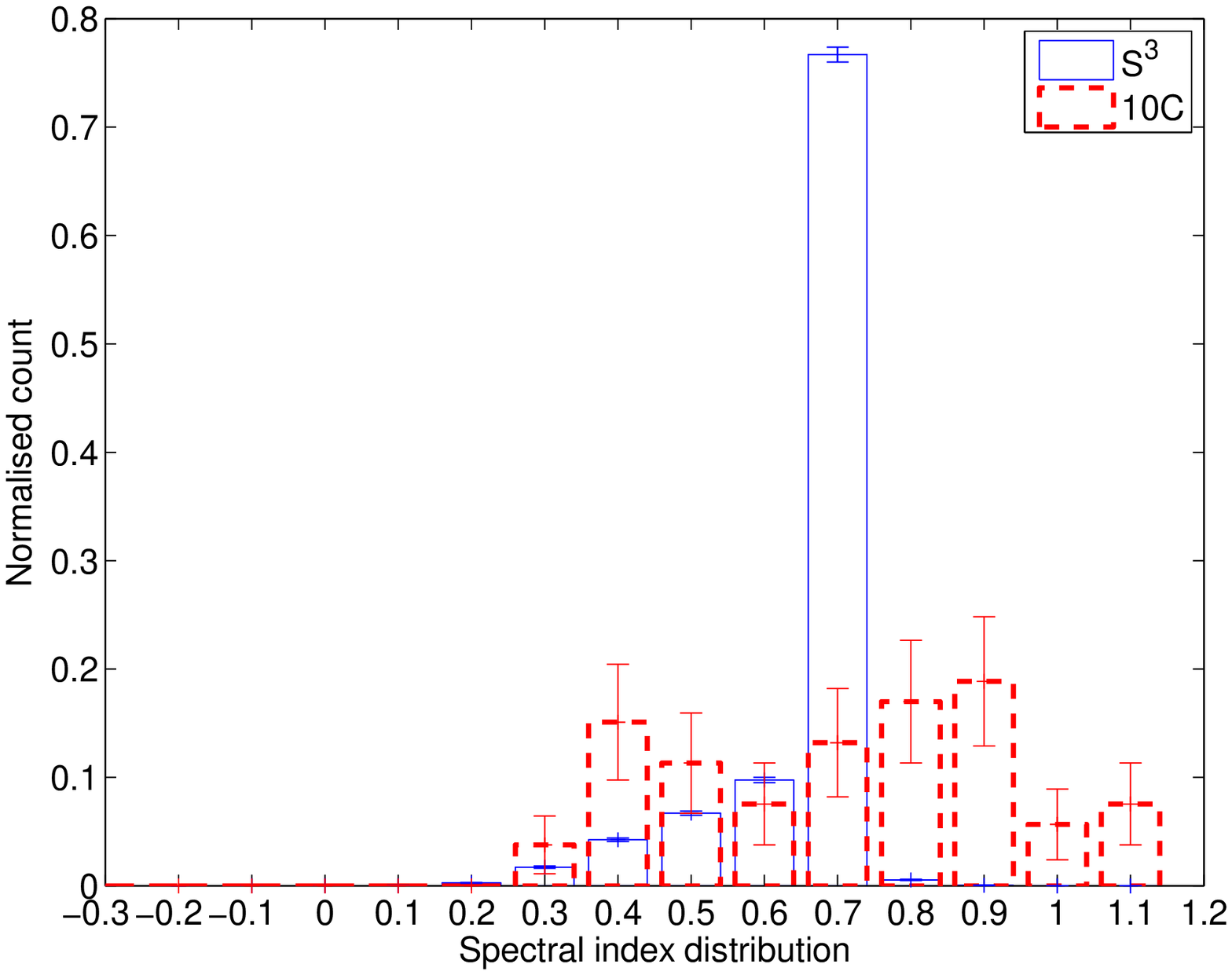}
            \includegraphics[width=8cm]{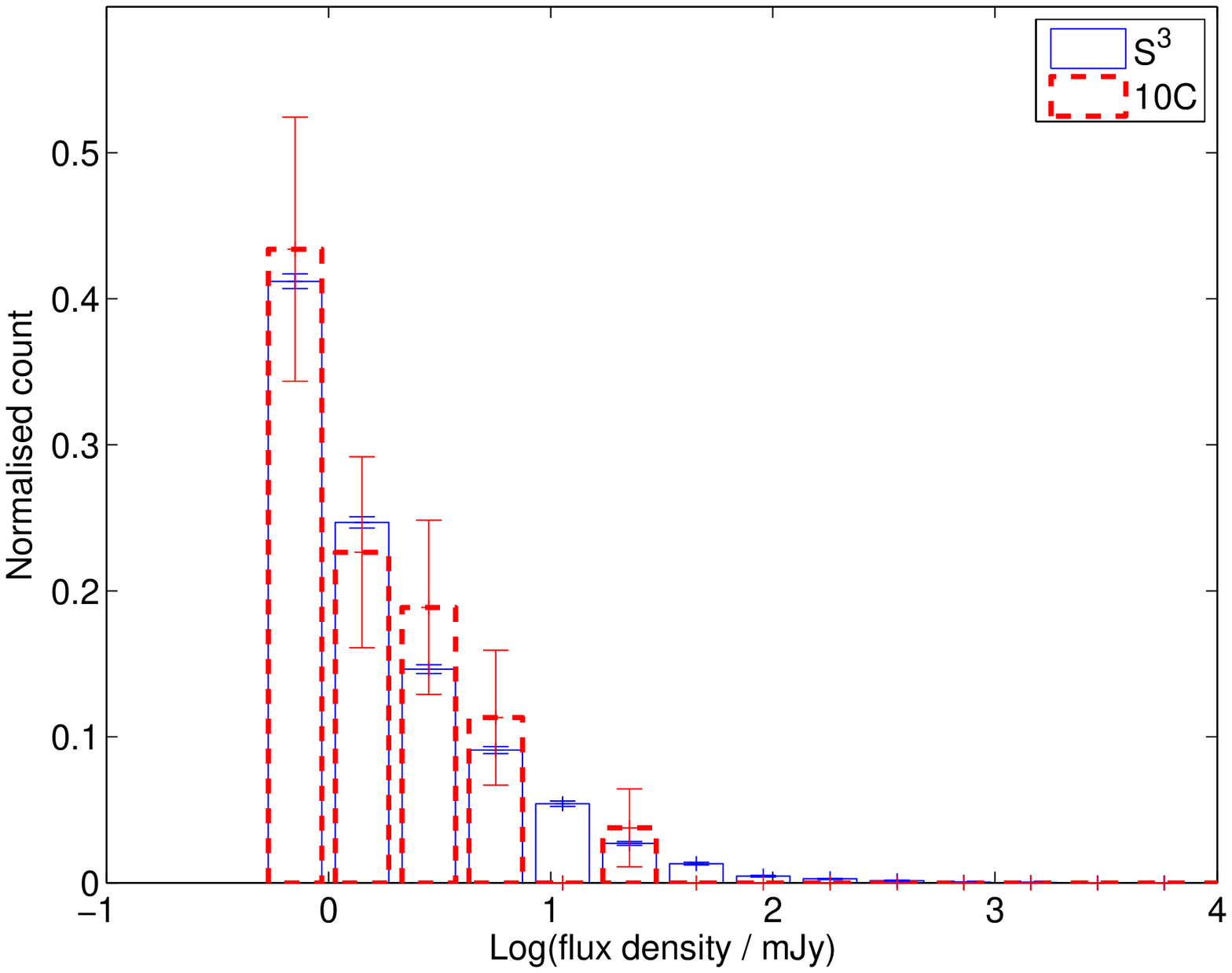}}
\caption{Comparison of the spectral index distribution and flux density distribution for sources in the $\rm S^3$ sample with $\rm S_{18~GHz} > 0.5~mJy$  and for 10C sources with $\rm S_{15.7~GHz} > 0.5~mJy$ and a spectral index ($\alpha^{15.7}_{0.61}$) greater than 0.3 only. Note that for the $\rm S^3$ sample the values plotted are $\alpha^{18}_{0.61}$ and 18-GHz flux density while for the 10C sources $\alpha^{15.7}_{0.61}$ and 15.7-GHz flux density are plotted. We do not expect this to make any difference to the results. Error bars show the poisson errors.}\label{fig:s-g-0.3}
\end{figure*}

The $\rm S^3$ sample is dominated by FRI sources, which make up 71 percent of the source population (Table \ref{tab:s-cubed}) in agreement with the discussion in Section \ref{section:counts}. The second largest source population are FRII sources, while radio quiet AGN, GPS sources and starforming galaxies each make only small contributions to the simulated source population. The $\rm S^3$ sample covers a region of area 400 $\rm deg^2$ while the deep 10C regions from which sample A is drawn cover 1.73 $\rm deg^2$. In order to investigate the possible impact of clustering on these results, five regions of the $\rm S^3$ sample with areas of 1.73 deg$^2$ were selected at random. The fraction of different sources types in each of these five regions was calculated and the range of values this gave is shown in Table \ref{tab:s-cubed}. Although the exact percentages of source vary across the five regions, the general proportions remain the same, indicating that clustering will not have a major effect on the 10C results.

The spectral index distributions of the different groups of sources in the simulation are shown in Fig. \ref{fig:s3-type-alpha}. All FRI, FRII and GPS sources have been modelled assuming their extended emission has a constant spectral index of 0.75, hence the prominent peaks at $\alpha = 0.75$. An orientation-dependent relativistic beaming model is used to find the contribution of the flat-spectrum core to the overall emission from each source; this gives the flatter spectrum tail in each spectral index distribution. The radio-quiet AGN have been assumed to have a constant spectral index of 0.7. The spectra of the starburst and starforming galaxies have been modelled using thermal and non-thermal components and in the case of starbursts a thermal dust component has been included.

\begin{table}
\caption{The proportion of different source types in the sub-sample of simulated sources selected to be directly comparable to the sources observed in this study. This sub-sample contains 16235 sources with a flux density at 18~GHz greater than 0.5~mJy. The range of percentages calculated from five regions of equivalent size to the 10C deep survey area is also given.}\label{tab:s-cubed}
\medskip
\begin{center}
\begin{tabular}{ldc}\hline
Source type           & \dhead{Percentage} & Range\\\hline
FRI                   & 71                 & 57 -- 85\\
FRII                  & 13                 & 6 -- 30\\
Radio quiet AGN       & 3                  & 0 -- 6\\
GPS                   & 3                  & 0 -- 8\\
Starburst             & 4                  & 0 -- 11\\
Quiescent starforming & 3                  & 1 -- 11\\\hline
\end{tabular}
\end{center}
\end{table}

Fig. \ref{fig:s-alpha-flux} shows a comparison of the spectral index and flux density distributions of the observed 10C sources (sample A, complete to 0.5~mJy) and $\rm S^3$ sample. It is clear that the simulation fails to reproduce the spectral index distribution of the 10C sources. The discrepancy for $\alpha > 0.7$ is due to the input assumption of the model that all sources have $\alpha = 0.75$; however the model fails to reproduce the distribution for $\alpha < 0.7$ with a conspicuous absence of sources with $\alpha < 0.3$. There are two main possibilities here; either the distribution of sources has not been modelled correctly and FRI sources do not dominate at this frequency and flux density level, but instead a new population with flat spectra is becoming important, or the emission from FRI sources has not been modelled correctly and that their flat spectrum cores are more dominant than predicted by the model in this frequency range. It is also possible that starburst galaxies may be causing this flattening in spectral index, although this is unlikely as it would require the contribution of starbursts to be greater than modelled by at least a factor of ten.

The actual and simulated flux density distributions are similar, but the 10C distribution contains a larger proportion of sources with flux densities less than 1~mJy. To test the possibility that there is a population of faint flat spectrum sources observed here which are missing from the simulation, the spectral index and flux density distributions were replotted, this time excluding all sources in the 10C sample with $\alpha < 0.3$ (Fig. \ref{fig:s-g-0.3}) as essentially no sources with $\alpha < 0.3$ are predicted by the model. The distributions are now more similar.

This analysis indicates that the extrapolations of the luminosity functions coupled with the models for the effects of beaming on the spectra of the radio-loud AGN used in this simulation have failed to reproduce the observed properties of the high frequency population. It is worth noting that the recent models of the source population at 15~GHz by \citet{2005A&A...431..893D} and \citet{2011A&A...533A..57T} also fail to reproduce the observed source count at flux densities $\lesssim$ 10~mJy, significantly under-predicting the observed number of sources. The updated version of the model of the 15-GHz source count by \citet{2005A&A...431..893D}, extracted from their website\footnote{http://web.oapd.inaf.it/rstools/srccnt\_tables}, shows that steep spectrum sources outnumber flat spectrum sources until the flux density drops below approximately 2 $\muup$Jy, in clear disagreement with the counts in Table \ref{tab:counts}. The \citeauthor{2005A&A...431..893D} model predicts that at $S_{15\rm{ GHz}} = 1$~mJy steep spectrum sources outnumber flat spectrum sources by nearly a factor of three. The results in Table \ref{tab:counts} show that for $S_{15\rm{ GHz}} < 1$~mJy there are twice as many flat spectrum sources as steep spectrum sources. The recent high frequency predictions of the source counts by \citet{2011A&A...533A..57T}  significantly under predict the number of sources observed at 15~GHz below approximately 5~mJy. This underprediction of the total number of sources could be explained by there being a greater number of flat spectrum sources at faint flux densities than are included in the model. These results highlight the difficulties inherent in predicting the behaviour of the high frequency radio source population by extrapolating from lower frequencies.

\section{Conclusions}\label{section:conclusions}

The radio spectral properties of 296 sources detected as part of the 10C survey at 15.7~GHz in the Lockman Hole are investigated in detail using a number of radio surveys, in particular a deep GMRT image at 610~MHz and a WSRT image at 1.4~GHz. Matches at other radio frequencies were found for 266 out of 296 sources, allowing their radio spectra to be investigated. For the 30 sources which were only detected at 15.7~GHz, upper limits are placed on their spectral indices.

There is a clear change in spectral index with flux density -- the median $\alpha^{15.7}_{0.61} = 0.75$ for flux densities greater than 1.5~mJy while the median $\alpha^{15.7}_{0.61} = 0.08$ for flux densities less than 0.8~mJy. This demonstrates that there is a population of flat spectrum sources emerging below 1~mJy. This result is consistent with results from other studies of the spectral indices of sources at lower frequencies.

The 10C source population was compared to two samples selected at 1.4~GHz from FIRST and NVSS at a comparable flux density. The spectral index distribution of these two samples is significantly different from that of the 10C sample selected at 15.7~GHz, the flat spectrum population present at 15~GHz being poorly represented in the 1.4-GHz samples. This demonstrates the well-known problem with extrapolating from lower frequencies to predict the properties of the high-frequency population.

The 10C sample was compared to a comparable sample selected from the SKADS Simulated Sky constructed by \citet{2008MNRAS.388.1335W}. The spectral index distributions of the two samples differ significantly; there are essentially no sources in the simulated sample with $\alpha < 0.3$ while 40 percent of the 10C sample have $\alpha^{15.7}_{0.61} < 0.3$. There is also a larger proportion of sources with flux densities below 1~mJy in the 10C sample than in the simulated sample -- 57 percent of the 10C sources have a flux density below 1~mJy compared to 40 percent of simulated sources. This indicates that the simulation does not accurately reproduce the observed population at 15.7~GHz. We conclude that either there is a population of faint, flat spectrum sources which are missing from the simulation or the high-frequency radio emission of a known population is not modelled correctly in the simulation. If the relative contributions of the different populations are modelled correctly, it is likely that the observed flat spectrum population is due to cores of FRI sources being much more dominant than the model suggests.

Our unique, faint 15~GHz samples are of great value when investigating the faint, high-frequency source population. We will use optical and infrared data in the Lockman Hole to investigate the nature of this population of sources in a later paper. We are also extending this study to fainter flux densities at 15~GHz.

\section*{Acknowledgements}

The authors thank Keith Grainge, Elizabeth Waldrem and Guy Pooley for useful discussions. IHW acknowledges an STFC studentship. We thank the anonymous referee for their helpful comments.

%
%

\setlength{\labelwidth}{0pt} 

\bsp

\label{lastpage}
\end{document}